\pdfoutput=1
\documentclass[11pt]{article}
\usepackage{footmisc}
\input{template_core/color_tokens}
\input{template_core/typography}
\input{template_core/layout}
\usepackage{microtype}
\usepackage{amsmath,amssymb,amsfonts,mathtools}
\usepackage{bm}
\usepackage{nicefrac}
\allowdisplaybreaks
\usepackage{graphicx}
\graphicspath{{figures/}{./}}
\usepackage{booktabs}
\usepackage{multirow,makecell,array}
\usepackage{subcaption}
\usepackage{algorithm}
\usepackage{algpseudocode}
\usepackage{enumitem}
\setlist[itemize]{leftmargin=2.2em,itemsep=2pt,topsep=2pt}
\setlist[enumerate]{leftmargin=2.2em,itemsep=2pt,topsep=2pt}
\usepackage{tikz}
\usetikzlibrary{positioning,calc,arrows.meta,shapes.geometric}
\usepackage[round,sort&compress]{natbib}
\usepackage{url}
\usepackage{hyperref}
\hypersetup{colorlinks=true,linkcolor=PaperLink,citecolor=PaperCitation,urlcolor=PaperUrl,breaklinks=true,bookmarksnumbered=true}
\usepackage{bookmark}
\usepackage[capitalise,nameinlink,noabbrev,sort&compress]{cleveref}
\numberwithin{equation}{section}

\providecommand{\PaperShortTitle}{Draft Paper}
\RequirePackage{titlesec}
\RequirePackage{fancyhdr}
\titleformat{\section}{\Large\PaperHeadingFont\bfseries\color{PaperAccentDark}}{\thesection}{0.7em}{}
\titleformat{\subsection}{\large\PaperHeadingFont\bfseries\color{PaperNavigation}}{\thesubsection}{0.7em}{}
\titleformat{\subsubsection}{\normalsize\PaperHeadingFont\bfseries\color{PaperNavigation}}{\thesubsubsection}{0.7em}{}
\titlespacing*{\section}{0pt}{3.0ex plus 0.5ex minus 0.2ex}{1.2ex}
\titlespacing*{\subsection}{0pt}{2.2ex plus 0.4ex minus 0.2ex}{0.8ex}
\titlespacing*{\subsubsection}{0pt}{1.6ex plus 0.3ex minus 0.2ex}{0.6ex}
\titleformat{\paragraph}[runin]{\normalsize\PaperBodyFont\bfseries\color{PaperInk}}{}{0pt}{}
\titlespacing*{\paragraph}{0pt}{1.4ex plus 0.4ex minus 0.2ex}{0.9em}
\RequirePackage[font=small,labelfont={bf,sf,color=PaperAccent},labelsep=colon]{caption}

\newcommand{\PaperHeaderSection}[1]{\markboth{}{#1}}
\newcommand{\PaperHeaderBaselineShift}{1pt}
\renewcommand{\headrulewidth}{0.45pt}
\renewcommand{\headrule}{\hbox to\headwidth{\color{PaperGold}\leaders\hrule height \headrulewidth\hfill}}
\fancypagestyle{plain}{\fancyhf{}\fancyfoot[C]{\small\PaperHeadingFont\color{PaperNavigation}\thepage}\renewcommand{\headrulewidth}{0pt}}

\usepackage{amsthm}
\usepackage{thm-restate}
\usepackage[most]{tcolorbox}
\usepackage{needspace}

\newcommand{\PaperTheoremHeadFont}{\PaperHeadingFont\bfseries}
\newcommand{\PaperTheoremBodyFont}{\itshape}
\newcommand{\PaperDefinitionHeadFont}{\PaperHeadingFont\bfseries}
\newcommand{\PaperDefinitionBodyFont}{\normalfont}
\newcommand{\PaperRemarkHeadFont}{\PaperHeadingFont\itshape}
\newcommand{\PaperRemarkBodyFont}{\normalfont}
\newtheoremstyle{paperplain}
  {0pt}{0pt}{\PaperTheoremBodyFont}{0pt}{\PaperTheoremHeadFont}{.}{0.5em}{}
\newtheoremstyle{paperdefinition}
  {0pt}{0pt}{\PaperDefinitionBodyFont}{0pt}{\PaperDefinitionHeadFont}{.}{0.5em}{}
\newtheoremstyle{paperremark}
  {0pt}{0pt}{\PaperRemarkBodyFont}{0pt}{\PaperRemarkHeadFont}{.}{0.5em}{}

\makeatletter
\newcommand{\PaperClearTheorem}[1]{%
  \expandafter\let\csname #1\endcsname\relax
  \expandafter\let\csname end#1\endcsname\relax
  \expandafter\let\csname c@#1\endcsname\relax
}
\makeatother

\theoremstyle{paperplain}
\PaperClearTheorem{theorem}
\newtheorem{theorem}{Theorem}[section]
\PaperClearTheorem{lemma}
\newtheorem{lemma}[theorem]{Lemma}
\PaperClearTheorem{proposition}
\newtheorem{proposition}[theorem]{Proposition}
\PaperClearTheorem{corollary}
\newtheorem{corollary}[theorem]{Corollary}
\PaperClearTheorem{conjecture}

\PaperClearTheorem{fact}
\newtheorem{fact}[theorem]{Fact}

\theoremstyle{paperdefinition}
\PaperClearTheorem{definition}
\newtheorem{definition}[theorem]{Definition}
\PaperClearTheorem{assumption}

\PaperClearTheorem{example}

\PaperClearTheorem{problem}

\PaperClearTheorem{openproblem}

\theoremstyle{paperremark}
\PaperClearTheorem{remark}
\newtheorem{remark}[theorem]{Remark}
\PaperClearTheorem{claim}

\makeatletter
\@for\PaperTheoremEnv:=lemma,proposition,corollary,conjecture,fact,definition,assumption,example,problem,openproblem,remark,claim\do{%
  \expandafter\def\csname the\PaperTheoremEnv\endcsname{\thesection.\arabic{theorem}}%
}
\makeatother

\newcommand{\PaperPlainBlockStyle}[1]{%
  \tcolorboxenvironment{#1}{
    enhanced jigsaw,
    boxrule=0pt,
    colframe=PaperAccent,
    colback=PaperAccentPale,
    arc=1.2mm,
    left=7pt,
    right=7pt,
    top=4pt,
    bottom=4pt,
    before=\Needspace{12\baselineskip},
    before skip=8pt,
    after skip=8pt
  }%
}
\newcommand{\PaperDefinitionBlockStyle}[1]{%
  \tcolorboxenvironment{#1}{
    enhanced jigsaw,
    boxrule=0pt,
    colframe=PaperDefinition,
    colback=PaperDefinitionLight,
    arc=1.2mm,
    left=7pt,
    right=7pt,
    top=4pt,
    bottom=4pt,
    before=\Needspace{12\baselineskip},
    before skip=8pt,
    after skip=8pt
  }%
}
\newcommand{\PaperRemarkBlockStyle}[1]{%
  \tcolorboxenvironment{#1}{
    enhanced jigsaw,
    boxrule=0pt,
    colframe=PaperRule,
    colback=white,
    arc=1.2mm,
    left=7pt,
    right=7pt,
    top=4pt,
    bottom=4pt,
    before=\Needspace{10\baselineskip},
    before skip=7pt,
    after skip=7pt
  }%
}

\floatstyle{ruled}
\restylefloat{algorithm}
\newcommand{\PaperAlgorithmLabelFont}{\bfseries}
\newcommand{\PaperAlgorithmCaptionFont}{\small\PaperBodyFont\color{PaperInk}}
\newcommand{\PaperAlgorithmBodyFont}{\small\PaperBodyFont\color{PaperInk}}
\newcommand{\PaperAlgorithmLineNumberFont}{\footnotesize\PaperBodyFont\color{PaperMutedInk}}
\newcommand{\PaperAlgorithmCommentFont}{\footnotesize\PaperBodyFont\itshape\color{PaperMutedInk}}
\newcommand{\PaperAlgorithmInputFont}{\PaperBodyFont\bfseries\color{PaperNavigation}}
\newcommand{\PaperAlgorithmOutputFont}{\PaperBodyFont\bfseries\color{PaperNavigation}}
\newcommand{\PaperAlgorithmKeywordFont}{\PaperBodyFont\bfseries\color{PaperInk}}
\newcommand{\PaperAlgorithmProcedureFont}{\PaperBodyFont\mdseries\color{PaperInk}}
\newcommand{\PaperAlgorithmInputText}{Input:}
\newcommand{\PaperAlgorithmOutputText}{Output:}
\DeclareCaptionFont{paperalgorithmlabel}{\PaperAlgorithmLabelFont}
\DeclareCaptionFont{paperalgorithmcaption}{\PaperAlgorithmCaptionFont}
\algrenewcommand{\algorithmicrequire}{\PaperAlgorithmInputFont\PaperAlgorithmInputText}
\algrenewcommand{\algorithmicensure}{\PaperAlgorithmOutputFont\PaperAlgorithmOutputText}
\algrenewcommand{\algorithmiccomment}[1]{\hfill{\PaperAlgorithmCommentFont\(\triangleright\)~#1}}
\algrenewcommand{\alglinenumber}[1]{{\PaperAlgorithmLineNumberFont #1}}
\algrenewcommand{\algorithmicif}{\PaperAlgorithmKeywordFont if}
\algrenewcommand{\algorithmicthen}{\PaperAlgorithmKeywordFont then}
\algrenewcommand{\algorithmicelse}{\PaperAlgorithmKeywordFont else}
\algrenewcommand{\algorithmicfor}{\PaperAlgorithmKeywordFont for}
\algrenewcommand{\algorithmicwhile}{\PaperAlgorithmKeywordFont while}
\algrenewcommand{\algorithmicdo}{\PaperAlgorithmKeywordFont do}
\algrenewcommand{\algorithmicend}{\PaperAlgorithmKeywordFont end}
\algrenewcommand{\algorithmicprocedure}{\PaperAlgorithmKeywordFont procedure}
\algrenewcommand{\textproc}[1]{{\PaperAlgorithmProcedureFont #1}}
\algrenewcommand{\algorithmicreturn}{\PaperAlgorithmKeywordFont return}
\AtBeginEnvironment{algorithmic}{\PaperAlgorithmBodyFont}
\newenvironment{paperalgorithmic}[1][1]{\begin{algorithmic}[#1]}{\end{algorithmic}}

\AtBeginDocument{%
  \PaperPlainBlockStyle{theorem}%
  \PaperPlainBlockStyle{lemma}%
  \PaperPlainBlockStyle{proposition}%
  \PaperPlainBlockStyle{corollary}%
  \PaperPlainBlockStyle{conjecture}%
  \PaperPlainBlockStyle{fact}%
  \PaperPlainBlockStyle{restatable}%
  \PaperDefinitionBlockStyle{definition}%
  \PaperDefinitionBlockStyle{assumption}%
  \PaperDefinitionBlockStyle{example}%
  \PaperDefinitionBlockStyle{problem}%
  \PaperDefinitionBlockStyle{openproblem}%
  \PaperRemarkBlockStyle{remark}%
  \PaperRemarkBlockStyle{claim}%
}

\crefname{theorem}{Theorem}{Theorems}\Crefname{theorem}{Theorem}{Theorems}
\crefname{lemma}{Lemma}{Lemmas}\Crefname{lemma}{Lemma}{Lemmas}
\crefname{proposition}{Proposition}{Propositions}\Crefname{proposition}{Proposition}{Propositions}
\crefname{corollary}{Corollary}{Corollaries}\Crefname{corollary}{Corollary}{Corollaries}
\crefname{conjecture}{Conjecture}{Conjectures}\Crefname{conjecture}{Conjecture}{Conjectures}
\crefname{fact}{Fact}{Facts}\Crefname{fact}{Fact}{Facts}
\crefname{definition}{Definition}{Definitions}\Crefname{definition}{Definition}{Definitions}
\crefname{assumption}{Assumption}{Assumptions}\Crefname{assumption}{Assumption}{Assumptions}
\crefname{example}{Example}{Examples}\Crefname{example}{Example}{Examples}
\crefname{problem}{Problem}{Problems}\Crefname{problem}{Problem}{Problems}
\crefname{openproblem}{Open Problem}{Open Problems}\Crefname{openproblem}{Open Problem}{Open Problems}
\crefname{remark}{Remark}{Remarks}\Crefname{remark}{Remark}{Remarks}
\crefname{claim}{Claim}{Claims}\Crefname{claim}{Claim}{Claims}
\crefname{algorithm}{Algorithm}{Algorithms}\Crefname{algorithm}{Algorithm}{Algorithms}

\DeclareMathOperator{\diag}{diag}
\DeclarePairedDelimiter{\abs}{\lvert}{\rvert}
\DeclarePairedDelimiter{\norm}{\lVert}{\rVert}
\DeclarePairedDelimiter{\set}{\{} {\}}
\DeclarePairedDelimiterX{\inner}[2]{\langle}{\rangle}{#1,#2}
\newcommand{\R}{\mathbb{R}}
\newcommand{\E}{\mathbb{E}}
\newcommand{\Prob}{\mathbb{P}}
\newcommand{\ind}{\mathbf{1}}
\newcommand{\T}{^{\mathsf{T}}}

\newcommand{\F}{\mathbb{F}_2^n}
\newcommand{\Ftuple}{(\mathbb{F}_2^n)^k}
\newcommand{\Tk}{T_k}
\newcommand{\phase}{\chi}
\newcommand{\Walsh}{\mathsf{Walsh}}
\newcommand{\Corr}{\mathsf{Corr}}
\newcommand{\Unif}{\mathrm{Unif}}
\newcommand{\supp}{\operatorname{supp}}
\newcommand{\TV}{\operatorname{TV}}
\newcommand{\Transcript}{\mathsf{Transcript}}
\newcommand{\Tr}{\operatorname{Tr}}
\newcommand{\rankF}{\operatorname{rank}_{\mathbb{F}_2}}
\newcommand{\cE}{\mathcal{E}}
\newcommand{\allones}{\mathbf{1}_n}
\newcommand{\affrank}{\operatorname{rk}}
\newcommand{\Per}{\operatorname{Per}}
\newcommand{\Meff}{M_{\mathrm{eff}}}
\newcommand{\Leff}{L_{\mathrm{eff}}}
\newcommand{\Ealign}{\cE_{\mathrm{align}}}
\newcommand{\Eodd}{\cE_{\mathrm{odd}}}

\renewcommand{\subsectionmark}[1]{}
\AtBeginDocument{%
}
\renewcommand{\PaperShortTitle}{Nonadaptive Cyclic-Forrelation Barriers}
\title{A Representation-Independent Second-Moment Barrier for Nonadaptive Cyclic-Forrelation Estimators}
\author{Mingda Wan, Deyuan Liu, and Yihao Hu}
\date{}
\hypersetup{pdftitle={A Representation-Independent Second-Moment Barrier for Nonadaptive Cyclic-Forrelation Estimators},pdfauthor={Mingda Wan, Deyuan Liu, Yihao Hu}}
\begin{document}
\thispagestyle{plain}
\begin{center}
{\LARGE\PaperHeadingFont\bfseries\color{PaperAccentDark}
A Representation-Independent Second-Moment Barrier for Nonadaptive Cyclic-Forrelation Estimators\par}
\vspace{0.65em}
{\normalsize\color{PaperInk}
\begin{tabular}{c}
Mingda Wan$^{*}$
\end{tabular}\par}
\vspace{0.40em}
{\normalsize\color{PaperInk}
\begin{tabular}{c@{\qquad}c}
Deyuan Liu$^{*}$ & Yihao Hu$^{*}$
\end{tabular}\par}
\vspace{0.20em}
{\small\color{PaperMutedInk}Westlake University\par}
\vspace{0.10em}
{\small\ttfamily\color{PaperMutedInk}\{wanmingda, liudeyuan, huyihao\}@westlake.edu.cn\par}
\vspace{0.40em}
{\color{PaperGold}\rule{\textwidth}{0.6pt}}
\end{center}
\vspace{0.25em}

\begin{center}
{\small\PaperHeadingFont\bfseries\color{PaperAccentDark} Abstract}
\end{center}
\vspace{-0.45em}
\begin{quote}
\small
Cyclic Forrelation asks for the normalized trace $\Tk$ of $k$ Boolean phase oracles on $\F$, alternating with $n$-qubit Hadamard transforms, where $N=2^n$. For every nonadaptive estimator with budgets $q_1,\ldots,q_k$, arbitrarily correlated query sets, and arbitrary nonlinear postprocessing, we prove $\E_{R,f}[G_R(f)^2]\ge N^{k-2}\Corr(h,\Tk)^2/\prod_j q_j$, with constant $1$ attained by Horvitz--Thompson Cartesian-product sampling. The proof projects each fixed transcript onto full-degree Walsh monomials and applies weighted Cauchy--Schwarz against an exact coverage budget. Thus the Cartesian scale $N^{1-2/k}$ is unavoidable when uniform target overlap and second moment are both bounded; unbiasedness alone gives no such restriction because a one-probe correction can restore it with heavy-tailed output. Separately, we construct a six-oracle complementary-subspace ensemble satisfying $W_6=\sigma I$ and an exact matching-certificate coupling, but give a certified adaptive algorithm recovering $\sigma$ in $n^{O(1)}N^{1/4}$ queries. These results delimit two approaches to the odd-cycle nonadaptive gap and the $\Omega(N^{1/4})$ adaptive lower bound; they do not close either gap for worst-case constant-error estimation or for a new hard distribution.

\end{quote}
\vspace{0.15em}
\begingroup
\renewcommand{\thefootnote}{\textasteriskcentered}
\footnotetext{Equal contribution.}
\endgroup
\section{Introduction}
\label{sec:introduction}

Forrelation is a canonical benchmark for exponential separations between quantum and classical query complexity. Introduced in the two-oracle setting by \citet{Aaronson2015Forrelation} and extended to $k$-fold alternating Fourier sums by \citet{Tal2020,bansal2021k}, standard $k$-Forrelation concerns an open chain of Boolean phase oracles. Cyclic Forrelation closes that chain into a trace. For integers $n,k\ge 1$, $N:=2^n$, and Boolean oracles $f_1,\ldots,f_k\colon\F\to\set{-1,+1}$, define
\begin{equation}
    \Tk(f_1,\ldots,f_k)
    :=\frac{1}{N}\Tr\!\left(
      D_{f_1}H^{\otimes n}D_{f_2}H^{\otimes n}\cdots D_{f_k}H^{\otimes n}
    \right),
    \label{eq:intro-trace}
\end{equation}
where $D_f$ is the diagonal matrix with entries $f(x)$ and $H$ is the single-qubit Hadamard gate. Because $\Tk$ is a normalized unitary trace, $k$ quantum oracle queries estimate it to constant additive error in the DQC1 model \citep{ji2026dqc1}. Its classical query landscape depends on the cycle and on adaptivity \citep{Shepherd2009,zhangshao2026cyclic}.

\subsection{Two gaps and the estimator-representation question}
\label{sub:intro-gap}

For fixed $k$ and worst-case constant additive error, the known classical query bounds split sharply by parity \citep{Shepherd2009,zhangshao2026cyclic}. For every even cycle $k=2m$, the nonadaptive lower bound $\Omega(N^{1-1/m})$ matches the Cartesian-product upper bound $O(N^{1-2/k})$. For $k=3$, \citet{zhangshao2026cyclic} give a biased nonadaptive Goldreich--Levin algorithm using $O(\log N\log\log N)$ queries, exponentially below the Cartesian scale $N^{1/3}$. For odd cycles $k=2m+1\ge 5$, a polynomial gap remains between $\Omega(N^{1-1/m})$ and $O(N^{1-2/k})$, while the best known adaptive lower bound is $\Omega(N^{1/4})$ for every $k\ge 4$.

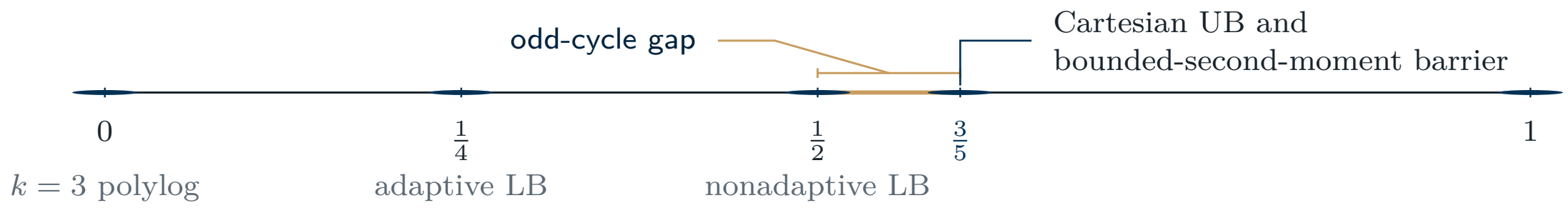
\begin{figure}[t]
\centering
\begin{tikzpicture}[x=1cm,y=1cm,>=Latex]
\begin{scope}[xscale=13.2]
  \draw[PaperInk,line width=0.55pt] (0,0) -- (1,0);
  \draw[PaperGold,line width=1.1pt] (0.50,0) -- (0.60,0);
  \foreach \x in {0,0.25,0.50,0.60,1} {
    \draw[PaperAccent,line width=0.45pt] (\x,0.045) -- (\x,-0.045);
    \fill[PaperAccent] (\x,0) circle (0.65pt);
  }
  \draw[PaperGold,line width=0.55pt] (0.50,0.18) -- (0.60,0.18);
  \draw[PaperGold,line width=0.55pt] (0.50,0.135) -- (0.50,0.225);
  \draw[PaperGold,line width=0.55pt] (0.60,0.135) -- (0.60,0.225);
  \draw[PaperGold,line width=0.55pt] (0.55,0.18) -- (0.47,0.48) -- (0.43,0.48);
  \node[anchor=east,font=\scriptsize\sffamily,text=PaperAccentDark] at (0.425,0.48) {odd-cycle gap};
  \draw[PaperAccent,line width=0.5pt] (0.60,0.065) -- (0.60,0.48) -- (0.65,0.48);
  \node[anchor=west,font=\scriptsize,align=left,text=PaperInk] at (0.655,0.48) {Cartesian UB and\\ bounded-second-moment barrier};
  \node[anchor=north,font=\scriptsize,text=PaperInk] at (0,-0.13) {$0$};
  \node[anchor=north,font=\scriptsize,text=PaperInk] at (0.25,-0.13) {$\tfrac14$};
  \node[anchor=north,font=\scriptsize,text=PaperInk] at (0.50,-0.13) {$\tfrac12$};
  \node[anchor=north,font=\scriptsize,text=PaperAccent] at (0.60,-0.13) {$\tfrac35$};
  \node[anchor=north,font=\scriptsize,text=PaperInk] at (1,-0.13) {$1$};
  \node[anchor=north,font=\scriptsize,text=PaperMutedInk] at (0,-0.62) {$k=3$ polylog};
  \node[anchor=north,font=\scriptsize,text=PaperMutedInk] at (0.25,-0.62) {adaptive LB};
  \node[anchor=north,font=\scriptsize,text=PaperMutedInk] at (0.50,-0.62) {nonadaptive LB};
\end{scope}
\end{tikzpicture}
\caption{Query exponents at the running example $k=5$ and the position of \Cref{thm:correlation-barrier}. Navy ticks mark known exponents, the gold segment marks the odd-cycle gap $[1/2,3/5]$, and the right callout marks the barrier for estimators with bounded second moment and constant uniform overlap; the $k=3$ algorithm of \citet{zhangshao2026cyclic} has exponent $0$.}
\label{fig:exponents}
\end{figure}

Our running nonadaptive example is $k=5$ ($m=2$), pictured in \Cref{fig:exponents}. The nonadaptive lower bound is $N^{1/2}$ and the Cartesian upper bound is $N^{3/5}$. The Cartesian estimator draws $q$ points per oracle and forms the Horvitz--Thompson average \citep{HorvitzThompson1952} of the full-degree Walsh terms in \eqref{eq:trace-expansion}. Its variance decomposes over agreement patterns between sampled tuple pairs. The full-overlap diagonal term is $N^{k-2}/q^k$: at $k=5$ it is $O(1)$ at $q=\Omega(N^{3/5})$ but grows as $N^{1/2}$ at $q=N^{1/2}$.

Two obstructions separate this variance computation from a query lower bound. An arbitrary nonadaptive estimator may correlate its query sets, reuse answers across coordinates, and apply arbitrary nonlinear postprocessing. Moreover, on uniform Boolean inputs $\norm{\Tk}_{2,\Unif}=1/N$, so a biased estimator can output nearly zero on almost every input while remaining accurate on high-trace instances. Exact unbiasedness alone is also not a barrier when heavy-tailed outputs are allowed (\Cref{prop:unbiased-tail}).

We therefore ask whether the full-overlap scale is forced for every nonadaptive estimator whose mean retains constant uniform overlap with $\Tk$ and whose output has bounded second moment. Independently, we test whether a two-shift complementary-subspace distribution on $k=6$, lifted to $k=7$ by a known quadratic cap, can push the adaptive lower bound beyond $\Omega(N^{1/4})$.

\subsection{Contributions and proof mechanisms}
\label{sub:intro-contributions}

Every contribution below maps to a labeled statement; \Cref{tab:scope-ledger} summarizes the hypotheses.
\begin{enumerate}
    \item \textbf{Representation-independent second-moment barrier (\Cref{thm:correlation-barrier,prop:tightness}).}
    For every nonadaptive estimator with budgets $q_1,\ldots,q_k>0$, arbitrarily correlated query sets, and arbitrary nonlinear postprocessing, if $h:=\E_R[G_R]$ and $\Corr(h,\Tk):=N^2\abs{\inner{h}{\Tk}_{\Unif}}$, then
    \[
        \E_{R,f}[G_R(f)^2]
        \ge\frac{N^{k-2}}{\prod_{j=1}^k q_j}\,\Corr(h,\Tk)^2.
    \]
    The proof also gives a general coverage-budget form, a total-query form, and an equality characterization. For integer budgets, Horvitz--Thompson Cartesian-product sampling attains constant $1$.

    \item \textbf{The $1/N$ threshold, unbiased-tail separation, and signed correlation (\Cref{thm:near-barrier,prop:unbiased-tail,cor:odd-scale,prop:hard-bridge}).}
    An $L_2(\Unif)$ radius $\eta$ contributes the factor $(1-N\eta)_+^2$, vanishing at the zero-estimator radius $\eta=1/N$. At $q_j\le N^{1-1/m}$ for $k=2m+1\ge 5$, the second moment is at least $\Corr(h,\Tk)^2N^{1/m}$. Conversely, a single random probe per oracle turns any bounded constant-error algorithm into an exactly unbiased one with error probability scaled by $1-p$ and second moment $\Omega(N^{k-2}/p)$. Finally, if a nonadaptive transcript is conditionally independent of a uniform sign $\sigma$ outside an event $\cE$, then $\abs{\E[\sigma Z]}\le 2\Prob[\cE]$ for every $Z\in[-1,1]$, with factor $1$ when $\cE\perp\sigma$.

    \item \textbf{Complementary-subspace ensemble and its adaptive ceiling (\Cref{prop:t6-identity,prop:t7-transfer,lem:matching-certificate,prop:posterior-predictive,prop:ensemble-upper}).}
    We construct a six-oracle ensemble whose alternating unitary is $\sigma I$, hence $T_6=\sigma$; the quadratic cap of \citet{zhangshao2026cyclic} gives $T_7=\sigma c_n$ with $\abs{c_n}\in\set{1/2,1/\sqrt2}$. Queries to $f_6$ form a matching in the quotient mask graph, so every adaptive algorithm has distinguishing advantage at most the probability of an aligned edge. If $f_6$ is deferred, that probability has an affine-rank posterior bound. Nevertheless, a certified adaptive algorithm recovers $\sigma$ in $n^{O(1)}N^{1/4}$ queries for every subspace $L$.
\end{enumerate}

\paragraph{Proof mechanisms.}
For \Cref{thm:correlation-barrier}, conditioning on $R$ expands the output in the Walsh basis of the queried bits. Nonadaptivity kills coefficients on uncovered tuples, pointwise Cauchy--Schwarz gives $\E_R[c_{R,x}^2]\ge\mu_x^2/\pi_x$, Parseval discards non-full characters, and weighted Cauchy--Schwarz applies the flat spectrum \eqref{eq:flat-spectrum} and coverage budget \eqref{eq:coverage-budget}. For the adaptive part, $L$- and $M$-periodicity collapse $H^{\otimes n}W_6H^{\otimes n}$ to $\widehat\phi*\widehat{\phi\psi}*\widehat\psi=\ind_0$, and the $f_6$-edges split the mask variables into independent two-variable blocks. The upper bound combines Goldreich--Levin filtering, direct period tests, birthday collisions, and a two-sided subspace certifier.

\subsection{Related work}
\label{sub:intro-related}

\paragraph{Standard versus cyclic Forrelation.}
\citet{Aaronson2015Forrelation} introduced two-oracle Forrelation to separate quantum and randomized query complexity; \citet{Tal2020,bansal2021k} established optimal randomized bounds for open-path $k$-Forrelation. Open-path Forrelation lacks the closing Hadamard trace, so its Fourier support and query-reuse structure differ from $\Tk$. In the cyclic setting, \citet{ji2026dqc1} established DQC1-completeness of normalized trace estimation for functions of log-local Hamiltonians, and \citet{MontanaroShao2024} studied query complexities of matrix functions more broadly.

\paragraph{Classical bounds for cyclic Forrelation.}
\citet{Shepherd2009} gave early lower bounds for alternating phase-Hadamard circuits, and \citet{zhangshao2026cyclic} established matching nonadaptive bounds for even cycles, the biased polylogarithmic $k=3$ algorithm, the odd-cycle gap, the adaptive $\Omega(N^{1/4})$ lower bound, and the quadratic-cap lift. Their variance analysis applies to linear Cartesian-product Horvitz--Thompson estimators \citep{HorvitzThompson1952}. \Cref{thm:correlation-barrier} removes linearity and product-grid independence in the bounded-second-moment, constant-overlap regime. \Cref{prop:unbiased-tail,prop:ensemble-upper} identify why the two tested routes do not close their gaps.

\paragraph{Fourier recovery and classical sampling.}
The adaptive upper bound uses heavy Fourier-coefficient recovery of \citet{GoldreichLevin1989} together with birthday collision search and differential period testing. The tightness witness and unbiased-tail construction use classical inverse-probability weighting \citep{HorvitzThompson1952}.

\paragraph{Organization.}
\Cref{sec:preliminaries} fixes the model and coverage budget. \Cref{sec:results} proves the nonadaptive results and states the unbiased-tail and signed-correlation propositions. \Cref{sec:complementary} develops the ensemble, matching coupling, deferred-$f_6$ bound, and adaptive upper bound. \Cref{sec:discussion} states the open problems. \Cref{app:proofs,app:ensemble-upper} contain the deferred proofs and the certified algorithm.

\section{Model and Walsh coverage}
\label{sec:preliminaries}

Fix integers $n,k\ge 1$, write $N:=\abs{\F}=2^n$, $[N]:=\set{1,\ldots,N}$, and $u_+:=\max\set{u,0}$. Let $f=(f_1,\ldots,f_k)$ range over $k$-tuples of Boolean functions $f_j\colon\F\to\set{-1,+1}$. Write $\Unif$ for the distribution in which all $kN$ oracle values are independent uniform signs, and define
\[
    \inner{u}{v}_{\Unif}:=\E_{f\sim\Unif}[u(f)v(f)],
    \qquad
    \norm{u}_{2,\Unif}:=\inner{u}{u}_{\Unif}^{1/2}.
\]
For $x,y\in\F$, write $\phase(x,y):=(-1)^{x\cdot y}$, where $x\cdot y=\sum_{r=1}^n x_r y_r\bmod 2$. Since $(H^{\otimes n})_{x,y}=N^{-1/2}\phase(x,y)$, expanding \eqref{eq:intro-trace} gives
\begin{equation}
    \Tk(f_1,\ldots,f_k)
    =N^{-k/2-1}\sum_{x=(x_1,\ldots,x_k)\in\Ftuple}
      \Walsh_x(f)\prod_{j=1}^k\phase(x_j,x_{j+1}),
    \qquad x_{k+1}:=x_1,
    \label{eq:trace-expansion}
\end{equation}
where $\Walsh_x(f):=\prod_{j=1}^k f_j(x_j)$. Because the oracles are distinct, the $N^k$ monomials $\set{\Walsh_x:x\in\Ftuple}$ are distinct Walsh characters and are orthonormal under $\inner{\cdot}{\cdot}_{\Unif}$, even when coordinates coincide. Setting $a_x:=N^{-k/2-1}\prod_{j=1}^k\phase(x_j,x_{j+1})$, the spectrum is flat:
\begin{equation}
    \abs{a_x}=N^{-k/2-1}
    \qquad\text{for every }x\in\Ftuple.
    \label{eq:flat-spectrum}
\end{equation}
Consequently (\Cref{app:normalization}),
\begin{equation}
    \norm{\Tk}_{2,\Unif}^2=N^k N^{-k-2}=N^{-2}.
    \label{eq:trace-norm}
\end{equation}
The cyclic phases determine only the signs of $a_x$; the nonadaptive barrier uses only \eqref{eq:flat-spectrum}.

\begin{definition}[Nonadaptive estimator]
\label{def:estimator}
Let $n,k\ge 1$ be integers and let $q_1,\ldots,q_k\in(0,\infty)$. A randomized estimator $G$ is \emph{nonadaptive} if its internal randomness $R$ is independent of $f$ and determines measurable query sets $Q_j(R)\subseteq\F$ before any oracle answer is observed, with $\abs{Q_j(R)}\le q_j$ almost surely. Its output $G_R(f)$ is measurable in $R$ for each fixed $f$ and depends only on $R$ and $\set{f_j(z):j\in[k],z\in Q_j(R)}$. Postprocessing randomness is absorbed into $R$. We assume $\E_R[\abs{G_R(f)}]<\infty$ for every Boolean input $f$, so that
\begin{equation}
    h(f):=\E_R[G_R(f)]
    \label{eq:mean-estimator}
\end{equation}
is well-defined. We use the mean \eqref{eq:mean-estimator} throughout the nonadaptive analysis.
\end{definition}

No independence is assumed among $Q_1(R),\ldots,Q_k(R)$, and conditional on $R$ the map from answers to output is arbitrary.

\begin{definition}[Normalized target overlap]
\label{def:normalized-overlap}
For integers $n,k\ge 1$ and any real-valued function $h$ on Boolean $k$-tuples, define
\begin{equation}
    \Corr(h,\Tk):=N^2\abs[\big]{\inner{h}{\Tk}_{\Unif}}.
    \label{eq:corr-def}
\end{equation}
\end{definition}

\begin{remark}[Interpretation of $\Corr(h,\Tk)$]
\label{rem:overlap}
By \eqref{eq:trace-norm}, $h=\Tk$ gives $\Corr(\Tk,\Tk)=1$, while $h=0$ gives $\Corr(0,\Tk)=0$. This is a rescaled inner product, not a cosine similarity, and can exceed $1$.
\end{remark}

\subsection{Walsh expansion of a fixed transcript}
\label{sub:walsh-interface}

Fix $R$. The full-degree Walsh coefficients of $G_R$ are
\[
    c_{R,x}:=\inner{G_R}{\Walsh_x}_{\Unif},\qquad x\in\Ftuple.
\]
If $x_j\notin Q_j(R)$ for some $j$, then $G_R$ does not depend on $f_j(x_j)$. Averaging over that bit gives
\begin{equation}
    c_{R,x}=0.
    \label{eq:uncovered-zero}
\end{equation}
Let $E_x:=\set{R:x_j\in Q_j(R)\text{ for every }j\in[k]}$, and define
\begin{equation}
    \pi_x:=\Prob_R[E_x],
    \qquad S_+:=\set{x\in\Ftuple:\pi_x>0}.
    \label{eq:coverage-prob}
\end{equation}
Summing $\ind_{E_x}(R)=\prod_{j=1}^k\ind_{x_j\in Q_j(R)}$ over $x$ gives the exact coverage budget
\begin{equation}
    \sum_{x\in S_+}\pi_x
    =\E_R\!\left[\prod_{j=1}^k\abs{Q_j(R)}\right]
    \le\prod_{j=1}^k q_j.
    \label{eq:coverage-budget}
\end{equation}
Nonadaptivity enters at \eqref{eq:uncovered-zero}; if a later query depended on earlier answers, averaging over an unqueried bit would be invalid. Since the input space is finite and $\E_R[\abs{G_R(f)}]<\infty$, the mean coefficients $\mu_x:=\E_R[c_{R,x}]$ are finite and vanish for $x\notin S_+$ by \eqref{eq:uncovered-zero}.

\section{Nonadaptive second-moment barrier}
\label{sec:results}

Combining \eqref{eq:flat-spectrum} with the coverage budget \eqref{eq:coverage-budget} gives a lower bound for arbitrary nonadaptive estimators.

\begin{theorem}[Correlation barrier]
\label{thm:correlation-barrier}
Let $n,k\ge 1$ be integers, $N:=2^n$, and let $G$ be a nonadaptive estimator for $\Tk$ (\Cref{def:estimator}) with mean $h:=\E_R[G_R]$, coverage probabilities $\pi_x$ from \eqref{eq:coverage-prob}, and support $S_+$.
\begin{itemize}
    \item \textbf{Part 1 (Second-moment barrier).} If $\sum_{x\in\Ftuple}\pi_x\le B$ for a real number $B>0$, then
    \begin{equation}
        \E_{R,f}[G_R(f)^2]
        \ge\frac{N^{k+2}}{B}\abs[\big]{\inner{h}{\Tk}_{\Unif}}^2
        =\frac{N^{k-2}}{B}\,\Corr(h,\Tk)^2.
        \label{eq:main-barrier}
    \end{equation}
    In particular, for budgets $q_1,\ldots,q_k>0$, taking $B=\prod_j q_j$ gives $\E_{R,f}[G_R(f)^2]\ge N^{k-2}\Corr(h,\Tk)^2/\prod_j q_j$.
    \item \textbf{Part 2 (Equality characterization).} Assume $\inner{h}{\Tk}_{\Unif}\neq0$ and $\E_{R,f}[G_R(f)^2]<\infty$. Equality in \eqref{eq:main-barrier} holds if and only if: (i) $\sum_{x\in S_+}\pi_x=B$; (ii) $G_R$ has zero Walsh energy outside $\set{\Walsh_x:x\in\Ftuple}$ almost surely; (iii) for every $x\in S_+$, $c_{R,x}$ is almost surely constant on $E_x$; and (iv) there is $\lambda\in\R\setminus\set{0}$ such that $\mu_x=\lambda\pi_x a_x$ for every $x\in S_+$.
\end{itemize}
\end{theorem}

\begin{proof}
For Part~1, assume $\E_{R,f}[G_R(f)^2]<\infty$; otherwise there is nothing to prove. Since the input space is finite and $R\perp f$, Fubini's theorem gives $\mu_x=\E_R[c_{R,x}]=\inner{h}{\Walsh_x}_{\Unif}$. By \eqref{eq:uncovered-zero}, $c_{R,x}=c_{R,x}\ind_{E_x}(R)$ and $\mu_x=0$ off $S_+$. If $S_+=\emptyset$, the right-hand side is zero. For $x\in S_+$, Cauchy--Schwarz over $R$ gives
\[
    \mu_x^2=\bigl(\E_R[c_{R,x}\ind_{E_x}(R)]\bigr)^2
    \le\E_R[c_{R,x}^2]\,\pi_x,
\]
so $\E_R[c_{R,x}^2]\ge\mu_x^2/\pi_x$. Parseval's identity for each fixed $R$ and discarding non-full characters yields
\[
    \E_{R,f}[G_R(f)^2]
    \ge\sum_{x\in S_+}\E_R[c_{R,x}^2]
    \ge\sum_{x\in S_+}\frac{\mu_x^2}{\pi_x}.
\]
Walsh orthogonality gives $\inner{h}{\Tk}_{\Unif}=\sum_{x\in S_+}\mu_x a_x$. Weighted Cauchy--Schwarz, \eqref{eq:flat-spectrum}, and $\sum_{x\in S_+}\pi_x\le B$ give
\begin{align*}
    \abs[\big]{\inner{h}{\Tk}_{\Unif}}^2
    &=\abs[\Bigg]{\sum_{x\in S_+}\frac{\mu_x}{\sqrt{\pi_x}}\cdot\sqrt{\pi_x}\,a_x}^2\\
    &\le\left(\sum_{x\in S_+}\frac{\mu_x^2}{\pi_x}\right)
      \left(\sum_{x\in S_+}\pi_x\abs{a_x}^2\right)
    \le\E_{R,f}[G_R(f)^2]\,N^{-k-2}B.
\end{align*}
Dividing by $N^{-k-2}B$ and using \eqref{eq:corr-def} gives \eqref{eq:main-barrier}. The per-oracle form follows from \eqref{eq:coverage-budget}.

For Part~2, $\inner{h}{\Tk}_{\Unif}\neq0$ makes $\sum_{x\in S_+}\mu_x^2/\pi_x>0$. Equality requires all four inequalities to be tight. Parseval's drop is tight exactly under~(ii); pointwise Cauchy--Schwarz is tight exactly under~(iii); weighted Cauchy--Schwarz is tight exactly when $\mu_x/\sqrt{\pi_x}=\lambda\sqrt{\pi_x}a_x$ for one $\lambda\neq0$, giving~(iv); and $\sum_x\pi_x\le B$ must be tight, giving~(i). Conversely, (i)--(iv) imply $c_{R,x}=\lambda a_x\ind_{E_x}(R)$ almost surely and $\E_{R,f}[G_R(f)^2]=\lambda^2N^{-k-2}B=(N^{k+2}/B)\abs{\inner{h}{\Tk}_{\Unif}}^2$. Without $\inner{h}{\Tk}_{\Unif}\neq0$, $G_R\equiv0$ gives $0=0$ without budget saturation.
\end{proof}

\begin{proposition}[Exact tightness and total-query form]
\label{prop:tightness}
Let $n,k\ge1$ be integers and $N:=2^n$.
\begin{itemize}
    \item \textbf{Part 1 (Exact tightness).} For positive integers $q_1,\ldots,q_k\in[N]$, equality in \eqref{eq:main-barrier} with $B=\prod_j q_j$ and $\Corr(h,\Tk)=1$ is attained by
    \[
        G_R^{\mathrm{HT}}(f):=
        \frac{N^k}{\prod_{j=1}^k q_j}
        \sum_{x\in Q_1(R)\times\cdots\times Q_k(R)}a_x\Walsh_x(f),
    \]
    where each $Q_j(R)$ is an independent uniform $q_j$-element subset of $\F$.
    \item \textbf{Part 2 (Total-query form).} For every $T>0$ and every nonadaptive estimator whose realized sets satisfy $\sum_{j=1}^k\abs{Q_j(R)}\le T$ almost surely,
    \[
        \E_{R,f}[G_R(f)^2]\ge\frac{k^kN^{k-2}}{T^k}\,\Corr(h,\Tk)^2.
    \]
\end{itemize}
\end{proposition}

\begin{proof}
For Part~1, $\pi_x=\prod_j(q_j/N)$ for every $x$, and $\sum_x\pi_x=\prod_jq_j$. The coefficient of $\Walsh_x$ is $(a_x/\pi_x)\ind_{E_x}(R)$, and all other coefficients vanish. Hence $\mu_x=a_x$, so $h=\Tk$ and $\Corr(h,\Tk)=1$. Parseval gives $\E_{R,f}[(G_R^{\mathrm{HT}})^2]=\sum_x\abs{a_x}^2/\pi_x=N^{k-2}/\prod_jq_j$.

For Part~2, AM--GM gives $\prod_j\abs{Q_j(R)}\le(T/k)^k$ pointwise. Taking expectations gives $\sum_x\pi_x\le(T/k)^k$, and Part~1 of \Cref{thm:correlation-barrier} with $B=(T/k)^k$ completes the proof. If $T<k$, some query set is empty and $\Corr(h,\Tk)=0$.
\end{proof}

\begin{remark}[Nonlinear postprocessing and correlated query sets]
\label{rem:nonlinear}
The proof never writes $G_R$ as a linear combination of sampled tuples. It projects an arbitrary function of the transcript onto the full-degree Walsh subspace.
\end{remark}

\subsection{Unbiased, \texorpdfstring{$L_2$}{L2}-near-unbiased, and odd-cycle consequences}
\label{sub:unbiased-consequences}

\begin{corollary}[Exact-unbiased barrier]
\label{cor:exact-barrier}
Let $G$ be a nonadaptive estimator with positive budgets $q_1,\ldots,q_k$. If $\E_R[G_R(f)]=\Tk(f)$ for every $f$, then $\E_{R,f}[G_R(f)^2]\ge N^{k-2}/\prod_jq_j$.
\end{corollary}

\begin{proof}
Set $h=\Tk$ in \Cref{thm:correlation-barrier} and use \eqref{eq:trace-norm}.
\end{proof}

\begin{theorem}[$L_2$-near-unbiased barrier]
\label{thm:near-barrier}
Let $\eta\ge0$ and let $G$ be a nonadaptive estimator with positive budgets $q_1,\ldots,q_k$ and mean $h:=\E_R[G_R]$. If $\norm{h-\Tk}_{2,\Unif}\le\eta$, then
\[
    \E_{R,f}[G_R(f)^2]
    \ge\frac{N^{k-2}}{\prod_{j=1}^k q_j}(1-N\eta)_+^2.
\]
\end{theorem}

\begin{proof}
Cauchy--Schwarz and \eqref{eq:trace-norm} give $\abs{\inner{h-\Tk}{\Tk}_{\Unif}}\le\eta/N$. Therefore
\[
    \inner{h}{\Tk}_{\Unif}
    =N^{-2}+\inner{h-\Tk}{\Tk}_{\Unif}
    \ge N^{-2}(1-N\eta).
\]
Thus $\Corr(h,\Tk)\ge(1-N\eta)_+$, and \Cref{thm:correlation-barrier} applies.
\end{proof}

\begin{remark}[Scope boundary: uniform overlap versus worst-case error]
\label{rem:uniform-vs-worstcase}
The factor $(1-N\eta)_+^2$ vanishes at $\eta=1/N$, the uniform error of the zero-query estimator. A constant-error estimator can therefore have $\Corr(h,\Tk)=o(1)$, as in the biased $k=3$ algorithm of \citet{zhangshao2026cyclic}. The barrier addresses constant overlap together with bounded second moment, not worst-case constant-error estimation alone.
\end{remark}

Exact unbiasedness alone imposes no query barrier when heavy-tailed outputs are allowed.

\begin{proposition}[Exact unbiasedness via heavy-tailed mixture]
\label{prop:unbiased-tail}
Let $n,k\ge1$, $N:=2^n$, and let $q>0$, $C\ge0$, $\varepsilon\ge0$, $\delta\in[0,1]$, and $p\in(0,1)$. Suppose $\tilde G_{R_1}$ is a nonadaptive estimator using at most $q$ queries per oracle, with $\abs{\tilde G_{R_1}(f)}\le C$ almost surely and $\Prob_{R_1}[\abs{\tilde G_{R_1}(f)-\Tk(f)}\le\varepsilon]\ge1-\delta$ for every $f$. Then there is a nonadaptive estimator $G_R^*$ using at most $q+1$ queries per oracle such that:
\begin{itemize}
    \item \textbf{Part 1 (Exact unbiasedness and error preservation).} For every $f$, $\E_R[G_R^*(f)]=\Tk(f)$ and $\Prob_R[\abs{G_R^*(f)-\Tk(f)}\le\varepsilon]\ge(1-p)(1-\delta)$.
    \item \textbf{Part 2 (Second-moment cost).}
    \[
        \E_{R,f}\bigl[G_R^*(f)^2\bigr]
        \ge\frac1p\max\!\left\{
          N^{k-2}-\frac{2(1-p)C}{N},\;
          \bigl(\sqrt{N^{k-2}}-(1-p)C\bigr)_+^2
        \right\}
        \ge\frac1p\bigl(\sqrt{N^{k-2}}-C\bigr)_+^2.
    \]
    In particular, if $C\le\frac14N^{k-1}$, then $\E_{R,f}[G_R^*(f)^2]\ge\frac12N^{k-2}/p$.
\end{itemize}
\end{proposition}

\begin{proof}
Draw $\xi\sim\mathrm{Bernoulli}(p)$ and $X\sim\Unif(\Ftuple)$ independently of $R_1$ and $f$, and query $Q_j(R_1)\cup\set{X_j}$ for each $j$. With probability $1-p$, output $\tilde G_{R_1}$. With probability $p$, output $p^{-1}(N^k a_X\Walsh_X(f)-(1-p)\tilde G_{R_1}(f))$. Since $\E_X[N^k a_X\Walsh_X(f)]=\Tk(f)$, the mean is exactly $\Tk(f)$. Restricting to $\xi=0$ gives error probability at least $(1-p)(1-\delta)$.

For the second moment, write $W_X:=N^k a_X\Walsh_X(f)$. Dropping the nonnegative $\xi=0$ term gives
\[
    \E_{R,f}[G_R^*(f)^2]
    \ge\frac1p\,\E_{R_1}\!\left[\E_{X,f}\bigl[(W_X-(1-p)\tilde G_{R_1})^2\bigr]\right].
\]
For fixed $R_1$, expansion gives
\begin{align*}
    &\E_{X,f}\bigl[(W_X-(1-p)\tilde G_{R_1})^2\bigr]\\
    &\qquad=\E_{X,f}[W_X^2]
      -2(1-p)\E_{X,f}[W_X\tilde G_{R_1}]
      +(1-p)^2\E_f[\tilde G_{R_1}^2].
\end{align*}
The first term is $N^{-k}\sum_xN^{2k}\abs{a_x}^2=N^{k-2}$. Independence and Cauchy--Schwarz give
\[
    \abs[\big]{\E_{X,f}[W_X\tilde G_{R_1}]}
    =\abs[\big]{\inner{\Tk}{\tilde G_{R_1}}_{\Unif}}
    \le\norm{\Tk}_{2,\Unif}\norm{\tilde G_{R_1}}_{2,\Unif}\le C/N.
\]
Discarding the nonnegative squared term yields the first lower bound. The reverse triangle inequality in $L_2(X,f)$ gives
\[
    \E_{X,f}\bigl[(W_X-(1-p)\tilde G_{R_1})^2\bigr]
    \ge\bigl(\sqrt{N^{k-2}}-(1-p)C\bigr)_+^2
    \ge\bigl(\sqrt{N^{k-2}}-C\bigr)_+^2.
\]
Taking expectations over $R_1$ proves both bounds. If $C\le\frac14N^{k-1}$, then $N^{k-2}-2(1-p)C/N\ge\frac12N^{k-2}$, giving the final claim.
\end{proof}

\begin{corollary}[Odd-cycle conjectured scale]
\label{cor:odd-scale}
Let $k=2m+1\ge5$ be fixed, and let $G$ be a nonadaptive estimator with positive budgets $q_1,\ldots,q_k$ and mean $h$.
\begin{itemize}
    \item \textbf{Part 1.} If $q_j\le N^{1-1/m}$ for every $j$, then $\E_{R,f}[G_R(f)^2]\ge\Corr(h,\Tk)^2N^{1/m}$.
    \item \textbf{Part 2.} As $n\to\infty$ with $k$ fixed, the joint conditions $\E_{R,f}[G_R(f)^2]=O(1)$ and $\Corr(h,\Tk)=\Omega(1)$ force $\prod_jq_j=\Omega(N^{k-2})$ and hence $\max_jq_j=\Omega(N^{1-2/k})$.
\end{itemize}
\end{corollary}

\begin{proof}
Part~1 follows by substitution: $N^{k-2}/\prod_jq_j\ge N^{(2m-1)-(2m+1)(1-1/m)}=N^{1/m}$. Part~2 rearranges \eqref{eq:main-barrier} to $\prod_jq_j\ge N^{k-2}\Corr^2/\E[G_R^2]=\Omega(N^{k-2})$, then takes the $k$-th root.
\end{proof}

At $k=5$, $q_j\le N^{1/2}$ forces second moment $\Corr(h,T_5)^2N^{1/2}$, whereas $\Corr=\Omega(1)$ and bounded second moment force $\max_jq_j=\Omega(N^{3/5})$.

\subsection{Correlation against a hidden sign outside a bad event}
\label{sub:hard-bridge}

Planted lower bounds often bound a collision probability, whereas estimator analyses use signed correlation. The next identity links the two with exact constants.

\begin{proposition}[Collision-gated correlation bound]
\label{prop:hard-bridge}
Let $(f,\sigma)$ be jointly distributed with $\sigma$ uniform on $\set{-1,+1}$, and let $\cE$ be an event. Let $R$ be independent of $(f,\sigma)$, with nonadaptive transcript $\tau=(Q(R),f|_{Q(R)})$ and output $Z=Z(\tau,R)\in[-1,1]$. Suppose $\tau$ is conditionally independent of $\sigma$ given $\neg\cE$ whenever $\Prob[\neg\cE]>0$.
\begin{itemize}
    \item \textbf{Part 1.} $\abs{\E[\sigma Z]}\le2\Prob[\cE]$.
    \item \textbf{Part 2.} If $\cE\perp\sigma$, then $\abs{\E[\sigma Z]}\le\Prob[\cE]$.
    \item \textbf{Part 3.} For every $p\in(0,1]$, a one-query Boolean oracle realizes $\Prob[\cE]=p/2$ and $\E[\sigma Z]=p=2\Prob[\cE]$.
\end{itemize}
\end{proposition}

\emph{Proof idea.} Each fiber $\set{\tau=(q_0,y_0)}=\set{Q(R)=q_0}\cap\set{f|_{q_0}=y_0}$ is rectangular, so $R\perp\sigma\mid\tau$ and $Z$ can be replaced by $\bar Z(\tau):=\E[Z\mid\tau]$. On $\neg\cE$, conditional independence and $\E[\sigma]=0$ allow subtracting $\gamma:=\E[\bar Z(\tau)\mid\neg\cE]$, yielding $\E[\sigma Z]=\E[\sigma(\bar Z(\tau)-\gamma)\ind_\cE]$. The absolute value is at most $2\Prob[\cE]$, or $\Prob[\cE]$ when $\E[\sigma\ind_{\neg\cE}]=0$. The full proof and tightness witness are in \Cref{app:deferred-nonadaptive}.

\begin{remark}[Scope boundary: externally supplied collision probabilities]
\label{rem:external-bridge}
\Cref{prop:hard-bridge} proves no collision bound of its own. It transfers any externally supplied bound on $\Prob[\cE]$ to signed correlation with factor at most $2$.
\end{remark}

\section{Complementary-subspace ensemble}
\label{sec:complementary}

We now test whether a structured distribution can improve the adaptive lower bound $\Omega(N^{1/4})$. The shortest even cycle with two independent shifts and two independent periodic masks is $k=6$, and the quadratic cap of \citet{zhangshao2026cyclic} lifts it to $k=7$. Our adaptive running example is the balanced case $\dim L=n/2$, so $s=t=N^{1/2}$.

\subsection{Complementary-subspace identity and \texorpdfstring{$T_7$}{T7} transfer}
\label{sub:t6-family}

Let $L\le\F$ and set $M:=L^\perp$. Let $\phi,\psi$ be $L$-periodic Boolean functions and let $A,B$ be $M$-periodic Boolean functions. For shifts $\alpha,\beta\in\F$ and $\sigma\in\set{-1,+1}$, write $\phase_\alpha(x):=(-1)^{\alpha\cdot x}$ and define
\begin{equation}
\begin{aligned}
    f_1(x)&=\phase_\alpha(x)\phi(x), & f_2(x)&=A(x),\\
    f_3(x)&=f_1(x)f_5(x), & f_4(x)&=B(x),\\
    f_5(x)&=\phase_\beta(x)\psi(x), & f_6(x)&=\sigma A(x+\alpha)B(x+\beta).
\end{aligned}
\label{eq:t6-family}
\end{equation}

\begin{definition}[Complementary-subspace and uniform-shift ensembles]
\label{def:cs-ensemble}
Fix $n\ge1$ and a prior over $(L,\alpha,\beta,\phi,\psi)$, where $L\le\F$, $\alpha,\beta\in\F$, and $\phi,\psi$ are $L$-periodic. Set $M:=L^\perp$. Conditional on these variables, draw $A,B\colon\F/M\to\set{-1,+1}$ independently and uniformly, and draw $\sigma$ independently and uniformly on $\set{-1,+1}$. The resulting law of \eqref{eq:t6-family} is a \emph{complementary-subspace ensemble}. It is the \emph{uniform-shift ensemble} when $(\alpha,\beta)$ is uniform on $\F\times\F$ and $\phi,\psi$ are independent uniform $L$-periodic functions.
\end{definition}

\begin{proposition}[Complementary-subspace identity]
\label{prop:t6-identity}
Let $n\ge1$. For every realization of \eqref{eq:t6-family}, the unitary
\[
    W_6:=D_{f_1}H^{\otimes n}D_{f_2}H^{\otimes n}\cdots D_{f_6}H^{\otimes n}
\]
satisfies $W_6=\sigma I$, and consequently $T_6(f_1,\ldots,f_6)=\sigma$.
\end{proposition}

\emph{Proof idea.} Conjugation by $H^{\otimes n}$ turns each odd-indexed oracle into a convolution supported on a translate of $M$. Since $A,B$ are $M$-periodic, the factors from $f_6$ cancel against $D_A,D_B$, leaving $\widehat\phi*\widehat{\phi\psi}*\widehat\psi=\widehat{\phi^2\psi^2}=\ind_0$. The full derivation is in \Cref{app:t6-proof}.

\begin{proposition}[Quadratic-cap transfer to $T_7$]
\label{prop:t7-transfer}
Let $n\ge1$ and define $Q(x):=(-1)^{\binom{\abs{x}_1}{2}}$, where $\abs{x}_1$ is Hamming weight. Set $\eta:=\allones$ if $n\equiv1\pmod4$ and $\eta:=0$ otherwise. For every six-tuple with $W_6=\sigma I$, the seven-tuple
\begin{equation}
    (g_1,\ldots,g_7):=(Qf_1,f_2,f_3,f_4,f_5,Qf_6,Q\phase_\eta)
    \label{eq:t7-transfer}
\end{equation}
satisfies $T_7(g_1,\ldots,g_7)=\sigma c_n$, where $c_n:=T_3(Q,Q,Q\phase_\eta)$ obeys $\abs{c_n}=1/2$ for even $n$ and $\abs{c_n}=1/\sqrt2$ for odd $n$.
\end{proposition}

\emph{Proof idea.} Right-multiplying $W_6=\sigma I$ by $H^{\otimes n}$ reduces the seven-cycle trace to $\sigma T_3(Q,Q,Q\phase_\eta)$. Writing $D_Q=\omega S^{\otimes n}+\bar\omega(S^\dagger)^{\otimes n}$ with $S=\diag(1,i)$ and $\omega=(1-i)/2$, all mixed single-qubit traces vanish, while $(SH)^3=e^{i\pi/4}I$ evaluates the two pure terms. The calculation is in \Cref{app:t6-proof}.

\begin{remark}[Attribution and query equivalence of the $T_7$ transfer]
\label{rem:t7-scope}
\Cref{prop:t7-transfer} specializes the quadratic cap of \citet{zhangshao2026cyclic}. Since $Q$ and $\phase_\eta$ are known, every query to $(g_1,\ldots,g_7)$ reduces to one query to $(f_1,\ldots,f_6)$.
\end{remark}

\subsection{Matching certificate, deferred \texorpdfstring{$f_6$}{f6} queries, and the adaptive ceiling}
\label{sub:adaptive-certificate}

Condition on $(L,\alpha,\beta)$ and write $[x]_M:=x+M\in\F/M$. The masks assign independent signs to the two parts $U=V=\F/M$, each of size $s:=\abs L=\abs{\F/M}$. A query $z$ to $f_6$ reads $\sigma A([z+\alpha]_M)B([z+\beta]_M)$ along
\[
    e(z):=\bigl([z+\alpha]_M,[z+\beta]_M\bigr).
\]
The right endpoint is determined by $v(u)=u+[\alpha+\beta]_M$, so distinct queried edges are vertex-disjoint and form a matching. Let $\Ealign$ be the event that some queried edge has both endpoints exposed by $f_2,f_4$, and let $\Eodd$ be the event that some odd set of queried edges has even degree at every unexposed vertex.

\begin{lemma}[Matching-certificate coupling]
\label{lem:matching-certificate}
Let $(f_1,\ldots,f_6)$ be drawn from a complementary-subspace ensemble. For any adaptive algorithm with randomness independent of the ensemble, let $\Transcript_+,\Transcript_-$ be its transcript laws under $\sigma=+1$ and $\sigma=-1$. Then $\Eodd=\Ealign$, and there is an online coupling agreeing up to the first query triggering $\Ealign$. In particular,
\[
    \TV(\Transcript_+,\Transcript_-)\le\Prob[\Ealign],
\]
where $\Prob[\Ealign]$ is the same under $\sigma=+1$, $\sigma=-1$, and the mixture.
\end{lemma}

\emph{Proof idea.} Every matching edge has degree $1$ at both endpoints, so $\Eodd=\Ealign$. Before alignment, each two-variable block has at most two of $\set{\tilde a_u,\tilde b_{v(u)},\tilde\sigma+\tilde a_u+\tilde b_{v(u)}}$ queried, whose mask components are independent. Every fresh coset answer is therefore uniform and independent of $\tilde\sigma$. The coupling stops immediately before reading the first aligned answer. The full proof is in \Cref{app:certificate-proof}.

\begin{figure}[!htbp]
\centering
\begin{tikzpicture}[x=1cm,y=1cm,>=Latex]
\node[font=\scriptsize\sffamily,text=PaperAccent] at (1.80,3.10) {left: $U=\F/M$};
\node[font=\scriptsize\sffamily,text=PaperNavigation] at (12.60,3.10) {right: $V=\F/M$};
\node[font=\scriptsize,text=PaperMutedInk,align=center] at (7.20,3.10) {queried $f_6$-edges form a matching};

\node[draw=PaperAccent,fill=PaperAccentLight,thick,circle,minimum size=0.64cm,inner sep=1pt,font=\scriptsize,text=PaperInk] (u1) at (1.80,2.20) {$u_1$};
\node[draw=PaperNavigation,fill=PaperNavigationLight,thick,circle,minimum size=0.64cm,inner sep=1pt,font=\scriptsize,text=PaperInk] (v1) at (12.60,2.20) {$v_1$};
\draw[very thick,PaperGold] (u1) -- (v1);
\node[font=\scriptsize,text=PaperAccentDark,align=center] at (7.20,2.52) {both ends exposed: $\tilde\sigma+\tilde a_{u_1}+\tilde b_{v_1}$ reveals $\sigma$};

\node[draw=PaperAccent,fill=PaperAccentLight,thick,circle,minimum size=0.64cm,inner sep=1pt,font=\scriptsize,text=PaperInk] (u2) at (1.80,1.10) {$u_2$};
\node[draw=PaperMutedInk,dashed,fill=white,circle,minimum size=0.64cm,inner sep=1pt,font=\scriptsize,text=PaperMutedInk] (v2) at (12.60,1.10) {$v_2$};
\draw[thick,PaperMutedInk] (u2) -- (v2);
\node[font=\scriptsize,text=PaperMutedInk] at (7.20,1.42) {$v_2$ hidden: fresh uniform sign (coupled across $\sigma=\pm1$)};

\node[draw=PaperMutedInk,dashed,fill=white,circle,minimum size=0.64cm,inner sep=1pt,font=\scriptsize,text=PaperMutedInk] (u3) at (1.80,0.00) {$u_3$};
\node[draw=PaperMutedInk,dashed,fill=white,circle,minimum size=0.64cm,inner sep=1pt,font=\scriptsize,text=PaperMutedInk] (v3) at (12.60,0.00) {$v_3$};
\draw[thick,PaperMutedInk] (u3) -- (v3);
\node[font=\scriptsize,text=PaperMutedInk] at (7.20,0.32) {neither end exposed: fresh uniform sign (coupled across $\sigma=\pm1$)};

\draw[PaperRule] (0.30,-0.60) -- (14.10,-0.60);
\node[draw=PaperAccent,fill=PaperAccentLight,thick,circle,minimum size=0.22cm,inner sep=0pt] at (0.55,-1.00) {};
\node[anchor=west,font=\scriptsize,text=PaperInk] at (0.76,-1.00) {exposed ($f_2$)};
\node[draw=PaperNavigation,fill=PaperNavigationLight,thick,circle,minimum size=0.22cm,inner sep=0pt] at (3.05,-1.00) {};
\node[anchor=west,font=\scriptsize,text=PaperInk] at (3.26,-1.00) {exposed ($f_4$)};
\node[draw=PaperMutedInk,dashed,fill=white,circle,minimum size=0.22cm,inner sep=0pt] at (5.55,-1.00) {};
\node[anchor=west,font=\scriptsize,text=PaperInk] at (5.76,-1.00) {hidden coset};
\draw[very thick,PaperGold] (8.05,-1.00) -- (8.65,-1.00);
\node[anchor=west,font=\scriptsize,text=PaperInk] at (8.76,-1.00) {aligned ($\Ealign$)};
\draw[thick,PaperMutedInk] (11.15,-1.00) -- (11.75,-1.00);
\node[anchor=west,font=\scriptsize,text=PaperInk] at (11.86,-1.00) {unaligned $f_6$-edge};
\end{tikzpicture}
\caption{Matching structure of queried $f_6$-edges in \Cref{lem:matching-certificate}, with $v_i=u_i+[\alpha+\beta]_M$. Shaded solid circles mark cosets exposed by $f_2$ or $f_4$, open dashed circles mark hidden cosets, muted lines mark unaligned queried edges, and the thick gold line marks the unique aligned configuration that reveals $\sigma$.}
\label{fig:certificate-app}
\end{figure}
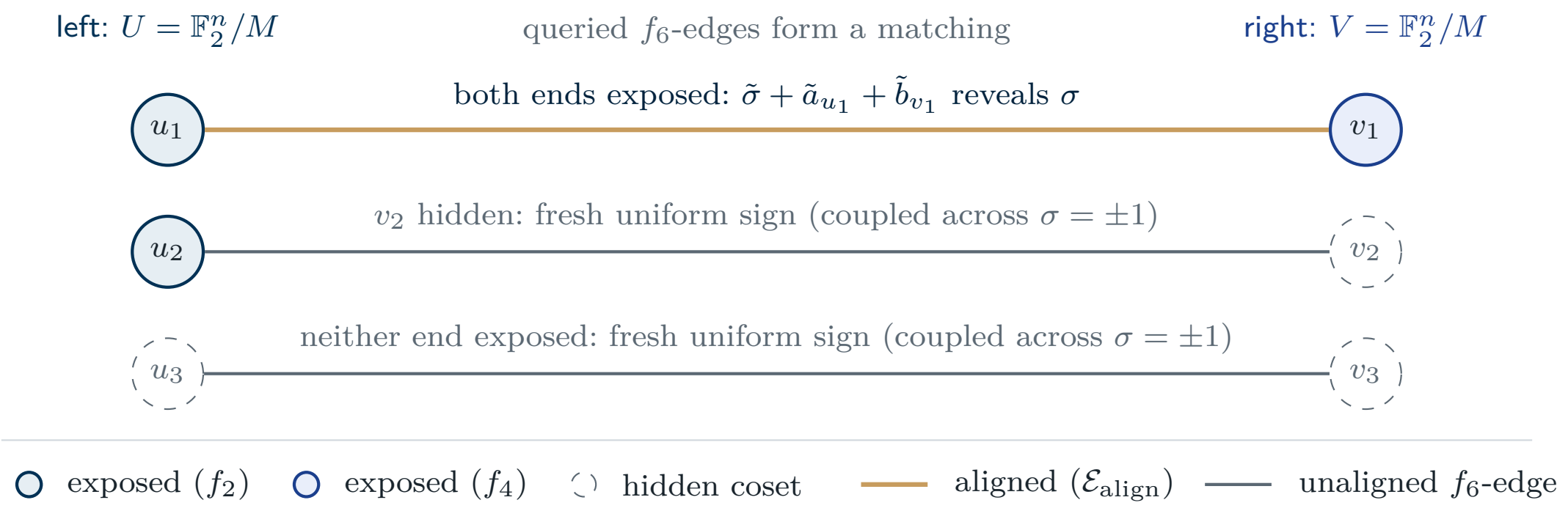

When $f_6$ answers are deferred, the alignment probability is controlled by the posterior peak of the quotient shifts.

\begin{proposition}[Deferred-$f_6$ alignment bound]
\label{prop:posterior-predictive}
Draw $(f_1,\ldots,f_6)$ from a complementary-subspace ensemble, and let $s:=\abs L=\abs{\F/M}$. Suppose an algorithm with independent internal randomness issues all queries to $f_1,\ldots,f_5$ and fixes $Q_6$ before reading any $f_6$ answer, with $\abs{Q_2}\le q_2$, $\abs{Q_4}\le q_4$, and $\abs{Q_6}\le q_6$ almost surely. Let $Y$ be the ordered shift-channel transcript, write $\bar\alpha:=[\alpha]_M$, $\bar\beta:=[\beta]_M$, and define $\rho(L,Y):=\max_{a,b\in\F/M}\Prob[(\bar\alpha,\bar\beta)=(a,b)\mid L,Y]$.
\begin{itemize}
    \item \textbf{Part 1.} $\Prob[\Ealign\mid L,Y]\le\min\set{1,q_2q_4q_6\rho(L,Y)}$.
    \item \textbf{Part 2.} On the uniform-shift ensemble, the posterior is uniform on an affine set of size $s^2/2^{\affrank(L,Y)}$, where $\affrank(L,Y):=\affrank(L,Q(Y))$ is the rank of \Cref{app:shift-normal-form}. Thus $\rho(L,Y)=2^{\affrank(L,Y)}/s^2$.
\end{itemize}
\end{proposition}

\emph{Proof idea.} Since $f_6$ is unread, the likelihood factors into an algorithm indicator, a mask factor independent of $(\alpha,\beta,\phi,\psi)$, and the shift-channel indicator. Bayes' rule gives the posterior given $(L,Y)$, and a union bound over $q_2q_4q_6$ triples proves Part~1. Eliminating $\phi,\psi$ over $\mathbb F_2$ leaves a consistent affine system of rank $\affrank(L,Y)$, proving Part~2. See \Cref{app:posterior-proof,app:shift-normal-form,app:posterior-rank}.

At the balanced example $s=t=N^{1/2}$, the baseline posterior peak is $1/s^2=1/N$. An adaptive algorithm can instead discover period subspaces and recover $\sigma$ directly.

\begin{proposition}[Adaptive recovery on the uniform-shift ensemble]
\label{prop:ensemble-upper}
For every integer $n\ge1$ and every subspace $L\le\F$, there is an adaptive algorithm (\Cref{alg:adaptive-cs}) making $n^{O(1)}N^{1/4}$ queries that outputs $\sigma$ with probability $1-o(1)$ as $n\to\infty$. Hence this ensemble cannot give a super-polylogarithmic, in particular polynomial $N^{1/4+\Omega(1)}$, improvement on $N^{1/4}$.
\end{proposition}

\emph{Proof idea.} Write $s=\abs L$ and $t=\abs M=N/s$. \Cref{alg:adaptive-cs} first runs Goldreich--Levin with empirical filtering on $f_1,f_5$. Every retained frequency has nonzero true coefficient and therefore lies in the correct coset; if $t\le256n^2$, Parseval guarantees such a frequency above the chosen threshold. Otherwise $t>256n^2$, so \Cref{lem:good-event} makes differential tests reject every $\ell\notin L$ with constant probability. Three branches cover the remaining cases: direct period sampling for $s\le64n^2$, birthday collisions on $f_2$ for $32n\le s\le N^{1/2}$, and differential birthday collisions on $f_1$ for $32n\le t\le N^{1/2}$. Before decoding, a universal verifier checks both $M'\subseteq\Meff$ by mask period tests and $M\subseteq M'$ by differential tests on $(M')^\perp$. These are precisely the hypotheses of \Cref{lem:eff-decode}, so an accepted branch returns $\sigma$ exactly. The full proof and query accounting are in \Cref{app:ensemble-upper}.

\begin{figure}[!htbp]
\centering
\begin{tikzpicture}[x=1cm,y=1cm,>=Latex]
\draw[PaperInk,thick] (0.80,0) -- (14.20,0);
\foreach \x in {0.80,2.60,4.50,10.50,12.40,14.20} {
  \draw[PaperAccent,thick] (\x,0.08) -- (\x,-0.08);
}
\node[diamond,draw=PaperGold,fill=PaperGoldLight,thick,inner sep=2.2pt] at (7.50,0) {};
\node[anchor=north,font=\scriptsize,align=center,text=PaperInk] at (0.80,-0.14) {$s=1$\\[-1pt] \textcolor{PaperMutedInk}{($t=N$)}};
\node[anchor=north,font=\scriptsize,text=PaperInk] at (2.60,-0.14) {$s=32n$};
\node[anchor=north,font=\scriptsize,text=PaperInk] at (4.50,-0.14) {$s=64n^2$};
\node[anchor=north,font=\scriptsize,align=center,text=PaperAccentDark] at (7.50,-0.18) {$s=t=N^{1/2}$\\[-1pt] \textcolor{PaperMutedInk}{(balanced example)}};
\node[anchor=north,font=\scriptsize,text=PaperInk] at (10.50,-0.14) {$t=256n^2$};
\node[anchor=north,font=\scriptsize,text=PaperInk] at (12.40,-0.14) {$t=32n$};
\node[anchor=north,font=\scriptsize,align=center,text=PaperInk] at (14.20,-0.14) {$s=N$\\[-1pt] \textcolor{PaperMutedInk}{($t=1$)}};

\draw[draw=PaperNavigation,fill=PaperNavigationLight,rounded corners=2pt,thick] (7.50,0.38) rectangle (12.40,1.04);
\node[font=\scriptsize,align=center,text=PaperInk] at (9.95,0.71) {\textbf{Branch~2:} diff.\ birthday on $f_1$\\[-1pt] $O(n^2N^{1/4}+n^8)$, certifies $L$};
\draw[draw=PaperAccent,fill=PaperAccentLight,rounded corners=2pt,thick] (2.60,1.22) rectangle (7.50,1.88);
\node[font=\scriptsize,align=center,text=PaperInk] at (5.05,1.55) {\textbf{Branch~1:} birthday on $f_2$\\[-1pt] $O(n^2N^{1/4}+n^8)$, certifies $M$};
\draw[draw=PaperDefinition,fill=PaperDefinitionLight,rounded corners=2pt,thick] (0.80,2.06) rectangle (4.50,2.72);
\node[font=\scriptsize,align=center,text=PaperInk] at (2.65,2.39) {\textbf{Branch~3:} direct periods\\[-1pt] $O(n^7)$, certifies $\Meff$};
\draw[draw=PaperGold,fill=PaperGoldLight,rounded corners=2pt,thick] (10.50,2.06) rectangle (14.20,2.72);
\node[font=\scriptsize,align=center,text=PaperAccentDark] at (12.35,2.39) {\textbf{Branch~4:} GL + filter\\[-1pt] $n^{O(1)}$, cosets $(w,w')$};

\draw[PaperAccent,thick] (0.80,2.92) -- (0.80,3.02) -- (10.40,3.02) -- (10.40,2.92);
\node[anchor=south,font=\scriptsize\sffamily,text=PaperAccentDark] at (5.60,3.06) {$t>256n^2$: universal certifier verifies $M\subseteq M'\subseteq\Meff$};
\draw[PaperGold,thick] (10.60,2.92) -- (10.60,3.02) -- (14.20,3.02) -- (14.20,2.92);
\node[anchor=south,font=\scriptsize\sffamily,text=PaperAccentDark] at (12.40,3.06) {self-certifying};
\end{tikzpicture}
\caption{Coverage of all $(s,t)$ with $st=N$ by \Cref{alg:adaptive-cs}; the horizontal axis is schematic, not a linear log scale. The gold bar marks the self-certifying Goldreich--Levin branch, while the green and blue bars mark branches checked by the universal verifier; the gold diamond marks the balanced example $s=t=N^{1/2}$.}
\label{fig:regimes}
\end{figure}
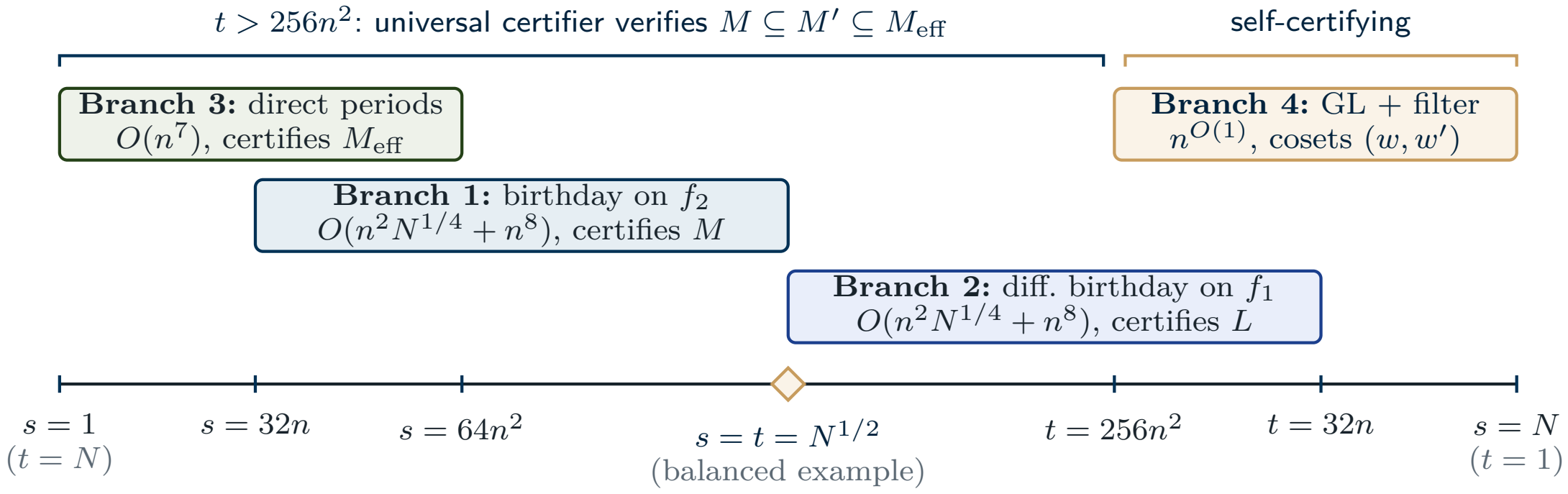

\begin{remark}[Scope boundary: deferred-$f_6$ and the $N^{1/4}$ ceiling]
\label{rem:adaptive-scope}
\Cref{lem:matching-certificate} applies to every adaptive algorithm, but \Cref{prop:posterior-predictive} requires $f_6$ to be deferred. Reading $f_6$ early invalidates the likelihood factorization. On this ensemble, \Cref{prop:ensemble-upper} matches $N^{1/4}$ up to $n^{O(1)}$ factors, so a larger exponent requires a different distribution.
\end{remark}

\section{Discussion and open problems}
\label{sec:discussion}

On the nonadaptive side, the full-overlap scale is intrinsic to uniform overlap under bounded second moment, independently of linearity, product-grid sampling, and query correlation. Beating $N^{1-2/k}$ requires either losing uniform overlap, as in the biased $k=3$ algorithm, or allowing heavy-tailed outputs. On the adaptive side, exact periodic cancellation exposes the hidden subspaces to period-finding and differential period-finding at the $N^{1/4}$ scale.

\begin{table}[!htbp]
\centering
{\scriptsize
\setlength{\tabcolsep}{3.5pt}
\begin{tabular}{@{}>{\raggedright\arraybackslash}p{0.21\linewidth}>{\raggedright\arraybackslash}p{0.17\linewidth}>{\raggedright\arraybackslash}p{0.31\linewidth}>{\raggedright\arraybackslash}p{0.25\linewidth}@{}}
\toprule
Result & Model & Conclusion & Scope boundary \\
\midrule
\Cref{thm:correlation-barrier,prop:tightness} & nonadaptive, arbitrary postprocessing & $\E[G_R^2]\ge N^{k-2}\Corr^2/B$; constant $1$ tight & requires $\Omega(1)$ uniform overlap; phases unused \\
\addlinespace
\Cref{cor:exact-barrier,thm:near-barrier,cor:odd-scale} & nonadaptive, uniform $L_2$ & factor $(1-N\eta)_+^2$; odd scale $N^{1/m}$ & vacuous at $\eta=1/N$; not worst-case constant error \\
\addlinespace
\Cref{prop:unbiased-tail} & nonadaptive mixture & exact unbiasedness in $q+1$ queries & second moment $\Omega(N^{k-2}/p)$ \\
\addlinespace
\Cref{prop:hard-bridge} & nonadaptive, $Z\in[-1,1]$ & $\abs{\E[\sigma Z]}\le2\Prob[\cE]$ (factor $1$ if $\cE\perp\sigma$) & collision probability supplied externally \\
\addlinespace
\Cref{prop:t6-identity,prop:t7-transfer} & exact operator identity & $W_6=\sigma I$, $T_7=\sigma c_n$ & quadratic cap from \citet{zhangshao2026cyclic}; same query cost \\
\addlinespace
\Cref{lem:matching-certificate} & every adaptive algorithm & $\TV\le\Prob[\Ealign]$ & does not bound alignment probability \\
\addlinespace
\Cref{prop:posterior-predictive} & adaptive, $f_6$ deferred & $\Prob[\Ealign\mid L,Y]\le\min\set{1,q_2q_4q_6\rho}$ & requires $f_6$ unread until other queries fixed \\
\addlinespace
\Cref{prop:ensemble-upper} & adaptive, uniform-shift & $n^{O(1)}N^{1/4}$ queries recover $\sigma$ & rules out super-polylog gain over $N^{1/4}$ \\
\bottomrule
\end{tabular}}
\caption{Summary of the results, models, and scope boundaries. The two nonadaptive boundaries concern bounded second moment and overlap; the adaptive bounds concern the particular ensemble of \Cref{def:cs-ensemble}.}
\label{tab:scope-ledger}
\end{table}

Three open problems remain adjacent to these boundaries:
\begin{enumerate}
    \item \textbf{Worst-case constant-error nonadaptive complexity for odd $k\ge5$.} Determine whether odd cycles admit a biased or heavy-tailed nonadaptive algorithm below $N^{1-2/k}$, or whether a worst-case lower bound exceeds $\Omega(N^{1-1/m})$.
    \item \textbf{Adaptive lower bounds above $\Omega(N^{1/4})$.} Construct a distribution with constant trace separation that resists mask period-finding and differential period-finding on the shift channel.
    \item \textbf{Planted non-uniform second-moment barriers.} Determine whether a distribution supported on high-trace tuples forces constant-error nonadaptive estimators to retain significant correlation with $\Tk$ under a weighted measure.
\end{enumerate}

\label{mainbody:end}
\clearpage
\PaperHeaderSection{References}
\bibliographystyle{plainnat}
\bibliography{refs}

\clearpage
\appendix
\section{Deferred proofs}
\label{app:proofs}

This appendix supplies the proofs deferred from \Cref{sec:preliminaries,sec:results,sec:complementary}. \Cref{app:normalization} verifies \eqref{eq:flat-spectrum} and \eqref{eq:trace-norm}. \Cref{app:deferred-nonadaptive} proves \Cref{prop:unbiased-tail,prop:hard-bridge}. \Cref{app:t6-proof} proves \Cref{prop:t6-identity,prop:t7-transfer}. \Cref{app:certificate-proof} proves \Cref{lem:matching-certificate}. \Cref{app:posterior-proof} proves Part~1 of \Cref{prop:posterior-predictive}, and \Cref{app:shift-normal-form,app:posterior-rank} prove Part~2. \Cref{app:ensemble-upper} contains the adaptive algorithm and proves \Cref{prop:ensemble-upper}.

\subsection{Coefficient normalization}
\label{app:normalization}

Because $(H^{\otimes n})_{u,v}=N^{-1/2}(-1)^{u\cdot v}$, expansion of \eqref{eq:intro-trace} gives
\[
    \Tk(f)=\sum_{x\in\Ftuple}a_x\Walsh_x(f),
    \qquad
    a_x=N^{-k/2-1}\prod_{j=1}^k\phase(x_j,x_{j+1}),
\]
where $x_{k+1}=x_1$. Each phase product lies in $\set{-1,+1}$, so $\abs{a_x}=N^{-k/2-1}$. The monomials $\Walsh_x$ are orthonormal under $\Unif$, hence
\[
    \norm{\Tk}_{2,\Unif}^2=\sum_{x\in\Ftuple}\abs{a_x}^2=N^kN^{-k-2}=N^{-2},
    \qquad
    \sum_{x\in\Ftuple}\abs{a_x}=N^{k/2-1}.
\]
The cyclic phases affect signs only and never these magnitudes.

\subsection{Deferred proofs for the unbiased-tail and conditioning propositions}
\label{app:deferred-nonadaptive}

\begin{proof}[Proof of \Cref{prop:unbiased-tail}]
Let $R_1$ be the randomness of $\tilde G$, and independently draw $\xi\sim\mathrm{Bernoulli}(p)$ and $X\sim\Unif(\Ftuple)$. Set $R:=(R_1,\xi,X)$ and $Q_j^*(R):=Q_j(R_1)\cup\set{X_j}$. Then $\abs{Q_j^*(R)}\le q+1$ almost surely. Write $W_X(f):=N^k a_X\Walsh_X(f)$ and define
\[
    G_R^*(f):=
    \begin{cases}
        \tilde G_{R_1}(f), & \xi=0,\\[3pt]
        p^{-1}\bigl(W_X(f)-(1-p)\tilde G_{R_1}(f)\bigr), & \xi=1.
    \end{cases}
\]
For every $f$, $G_R^*(f)$ is bounded, hence integrable.

\textbf{Part 1.} Uniformity of $X$ and \eqref{eq:trace-expansion} give $\E_X[W_X(f)]=\Tk(f)$. By mutual independence,
\begin{align*}
    \E_R[G_R^*(f)]
    &=(1-p)\E_{R_1}[\tilde G_{R_1}(f)]
      +\E_X[W_X(f)]-(1-p)\E_{R_1}[\tilde G_{R_1}(f)]
    =\Tk(f).
\end{align*}
Thus $\Corr(h^*,\Tk)=1$ by \eqref{eq:trace-norm}. Restricting to $\xi=0$ and using $\xi\perp R_1$ gives
\[
    \Prob_R[\abs{G_R^*(f)-\Tk(f)}\le\varepsilon]
    \ge(1-p)\Prob_{R_1}[\abs{\tilde G_{R_1}(f)-\Tk(f)}\le\varepsilon]
    \ge(1-p)(1-\delta).
\]

\textbf{Part 2.} Dropping the nonnegative $\xi=0$ term gives
\[
    \E_{R,f}[G_R^*(f)^2]
    \ge\frac1p\,\E_{R_1}\!\left[\E_{X,f}\bigl[(W_X(f)-(1-p)\tilde G_{R_1}(f))^2\bigr]\right].
\]
Fix $R_1$ with $\abs{\tilde G_{R_1}(f)}\le C$ for all $f$. Expanding the square gives
\begin{align*}
    &\E_{X,f}\bigl[(W_X(f)-(1-p)\tilde G_{R_1}(f))^2\bigr]\\
    &\qquad=\E_{X,f}[W_X(f)^2]
      -2(1-p)\E_{X,f}[W_X(f)\tilde G_{R_1}(f)]
      +(1-p)^2\E_f[\tilde G_{R_1}(f)^2].
\end{align*}
Since $\Walsh_x(f)^2=1$ and $\abs{a_x}^2=N^{-k-2}$,
\[
    \E_{X,f}[W_X(f)^2]
    =N^{-k}\sum_{x\in\Ftuple}N^{2k}\abs{a_x}^2=N^{k-2}.
\]
Because $X\perp(R_1,f)$ and $\E_X[W_X(f)]=\Tk(f)$, Cauchy--Schwarz gives
\[
    \abs[\big]{\E_{X,f}[W_X(f)\tilde G_{R_1}(f)]}
    =\abs[\big]{\inner{\Tk}{\tilde G_{R_1}}_{\Unif}}
    \le\norm{\Tk}_{2,\Unif}\norm{\tilde G_{R_1}}_{2,\Unif}\le C/N.
\]
Discarding the nonnegative final term therefore yields
\[
    \E_{X,f}\bigl[(W_X-(1-p)\tilde G_{R_1})^2\bigr]
    \ge N^{k-2}-\frac{2(1-p)C}{N}.
\]
Independently, the reverse triangle inequality in $L_2(X,f)$ gives
\[
    \E_{X,f}\bigl[(W_X-(1-p)\tilde G_{R_1})^2\bigr]
    \ge\bigl(\sqrt{N^{k-2}}-(1-p)C\bigr)_+^2
    \ge\bigl(\sqrt{N^{k-2}}-C\bigr)_+^2.
\]
Taking expectations over $R_1$ and multiplying by $1/p$ proves both displayed bounds. Finally, if $C\le\frac14N^{k-1}$,
\[
    N^{k-2}-\frac{2(1-p)C}{N}
    \ge N^{k-2}-\frac{2\cdot\frac14N^{k-1}}N
    =\frac12N^{k-2},
\]
so $\E_{R,f}[G_R^*(f)^2]\ge\frac12N^{k-2}/p$.
\end{proof}

\begin{proof}[Proof of \Cref{prop:hard-bridge}]
For every realized transcript $(q_0,y_0)$ of positive probability, the fiber is rectangular:
\[
    \set{\tau=(q_0,y_0)}
    =\set{Q(R)=q_0}\cap\set{f|_{q_0}=y_0}.
\]
The first event depends only on $R$ and the second only on $f$. For every $R$-event $A$ and $s\in\set{-1,+1}$, independence of $R$ and $(f,\sigma)$ gives
\[
    \Prob[R\in A,\sigma=s,\tau=(q_0,y_0)]
    =\Prob[R\in A,Q(R)=q_0]\,\Prob[\sigma=s,f|_{q_0}=y_0].
\]
Dividing by $\Prob[\tau=(q_0,y_0)]>0$ shows $R\perp\sigma\mid\tau$. Set $\bar Z(\tau):=\E[Z(\tau,R)\mid\tau]\in[-1,1]$. The tower property gives
\[
    \E[\sigma Z]
    =\E\bigl[\E[\sigma Z\mid\tau]\bigr]
    =\E\bigl[\E[\sigma\mid\tau]\E[Z\mid\tau]\bigr]
    =\E[\sigma\bar Z(\tau)].
\]

\textbf{Parts 1 and 2.} If $\Prob[\cE]=1$, both bounds hold trivially. Otherwise let $\gamma:=\E[\bar Z(\tau)\mid\neg\cE]\in[-1,1]$. Since $\tau\perp\sigma\mid\neg\cE$,
\[
    \E[\sigma\bar Z(\tau)\ind_{\neg\cE}]
    =\Prob[\neg\cE]\,\E[\sigma\bar Z(\tau)\mid\neg\cE]
    =\gamma\,\E[\sigma\ind_{\neg\cE}].
\]
Because $\E[\sigma]=0$, $\E[\sigma\ind_{\neg\cE}]=-\E[\sigma\ind_\cE]$, hence
\[
    \E[\sigma Z]
    =\E\bigl[\sigma(\bar Z(\tau)-\gamma)\ind_\cE\bigr].
\]
Since $\abs{\bar Z(\tau)-\gamma}\le2$, Part~1 follows, including $\Prob[\cE]=0$. If $\cE\perp\sigma$, then $\E[\sigma\ind_{\neg\cE}]=0$, so $\E[\sigma Z]=\E[\sigma\bar Z(\tau)\ind_\cE]$ and $\abs{\E[\sigma Z]}\le\Prob[\cE]$.

\textbf{Part 3.} Query only $f_1(0)$ and output $Z:=f_1(0)$. Given $\sigma=+1$, set $f_1(0)=+1$ with probability $p$ and $-1$ otherwise; given $\sigma=-1$, set $f_1(0)=-1$. Let $\cE:=\set{f_1(0)=+1}$. Then $\Prob[\cE]=p/2$, and on $\neg\cE$ the transcript is deterministic, so $\tau\perp\sigma\mid\neg\cE$. Finally,
\[
    \E[\sigma Z]
    =\frac12\bigl(p-(1-p)\bigr)+\frac12=p=2\Prob[\cE].
\]
\end{proof}

\subsection{Proofs of the complementary-subspace identity and the \texorpdfstring{$T_7$}{T7} transfer}
\label{app:t6-proof}

\begin{proof}[Proof of \Cref{prop:t6-identity}]
For $g\colon\F\to\R$, define $\widehat g(\xi):=N^{-1}\sum_{y\in\F}g(y)(-1)^{\xi\cdot y}$. For every $u,v\in\F$,
\[
    (H^{\otimes n}D_gH^{\otimes n})_{u,v}
    =\frac1N\sum_{y\in\F}(-1)^{u\cdot y}g(y)(-1)^{y\cdot v}
    =\widehat g(u+v),
\]
so $(H^{\otimes n}D_gH^{\otimes n})\lvert v\rangle=\sum_{\xi\in\F}\widehat g(\xi)\lvert v+\xi\rangle$. If $g$ is $L$-periodic, changing $y\mapsto y+\ell$ gives $\widehat g(\xi)=(-1)^{\xi\cdot\ell}\widehat g(\xi)$ for every $\ell\in L$, hence $\widehat g$ is supported on $M=L^\perp$. Applying this to $\phi,\psi,\phi\psi$ yields
\[
    \widehat{f_1}(\xi)=\widehat\phi(\xi+\alpha),\qquad
    \widehat{f_3}(\xi)=\widehat{\phi\psi}(\xi+\alpha+\beta),\qquad
    \widehat{f_5}(\xi)=\widehat\psi(\xi+\beta).
\]
Since $H^{\otimes n}$ is an involution,
\[
    H^{\otimes n}W_6H^{\otimes n}
    =(H^{\otimes n}D_{f_1}H^{\otimes n})D_{f_2}
     (H^{\otimes n}D_{f_3}H^{\otimes n})D_{f_4}
     (H^{\otimes n}D_{f_5}H^{\otimes n})D_{f_6}.
\]
Fix $x$ and apply the factors from right to left. First,
\[
    (H^{\otimes n}D_{f_5}H^{\otimes n})D_{f_6}\lvert x\rangle
    =\sigma A(x+\alpha)B(x+\beta)\sum_{v\in M}\widehat\psi(v)\lvert x+\beta+v\rangle.
\]
Because $B$ is $M$-periodic and $v\in M$, applying $D_B$ cancels $B(x+\beta)$ via $B^2=1$. Next,
\begin{align*}
    &(H^{\otimes n}D_{f_3}H^{\otimes n})D_{f_4}
      (H^{\otimes n}D_{f_5}H^{\otimes n})D_{f_6}\lvert x\rangle\\
    &\qquad=\sigma A(x+\alpha)
      \sum_{v,r\in M}\widehat\psi(v)\widehat{\phi\psi}(r)
      \lvert x+\alpha+v+r\rangle.
\end{align*}
Since $A$ is $M$-periodic and $v+r\in M$, applying $D_A$ cancels $A(x+\alpha)$. Finally, shifting by $\alpha+u$ for $u\in M$ gives
\[
    (H^{\otimes n}W_6H^{\otimes n})\lvert x\rangle
    =\sigma\sum_{z\in M}
      (\widehat\phi*\widehat{\phi\psi}*\widehat\psi)(z)\lvert x+z\rangle.
\]
With the present normalization, $g(y)=\sum_\xi\widehat g(\xi)(-1)^{\xi\cdot y}$, so $\widehat{gh}=\widehat g*\widehat h$ without a factor of $N$. Therefore
\[
    (\widehat\phi*\widehat{\phi\psi}*\widehat\psi)(z)
    =\widehat{\phi\cdot(\phi\psi)\cdot\psi}(z)
    =\widehat{\phi^2\psi^2}(z)
    =\widehat{\ind}(z)
    =\ind_{z=0}.
\]
Hence $H^{\otimes n}W_6H^{\otimes n}=\sigma I$, so $W_6=\sigma I$ and $T_6=N^{-1}\Tr(W_6)=\sigma$.
\end{proof}

\begin{proof}[Proof of \Cref{prop:t7-transfer}]
Right-multiplying $W_6=\sigma I$ by $H^{\otimes n}$ gives
\[
    D_{f_1}H^{\otimes n}\cdots D_{f_6}=\sigma H^{\otimes n}.
\]
Diagonal matrices commute, so $D_{Qf_1}=D_QD_{f_1}$ and $D_{Qf_6}=D_{f_6}D_Q$. Substituting \eqref{eq:t7-transfer} into \eqref{eq:intro-trace} yields
\begin{align*}
    T_7(g_1,\ldots,g_7)
    &=\frac1N\Tr\!\left(D_Q(D_{f_1}H^{\otimes n}\cdots D_{f_6})D_QH^{\otimes n}D_{Q\phase_\eta}H^{\otimes n}\right)\\
    &=\frac\sigma N\Tr\!\left(D_QH^{\otimes n}D_QH^{\otimes n}D_{Q\phase_\eta}H^{\otimes n}\right)
    =\sigma T_3(Q,Q,Q\phase_\eta).
\end{align*}
Set $S:=\diag(1,i)$ and $\omega:=(1-i)/2=e^{-i\pi/4}/\sqrt2$. Checking $w\bmod4$ gives
\[
    (-1)^{\binom w2}=\omega i^w+\bar\omega(-i)^w,
    \qquad
    (-1)^{\binom w2+w}=\bar\omega i^w+\omega(-i)^w.
\]
Consequently,
\[
    D_Q=\omega S^{\otimes n}+\bar\omega(S^\dagger)^{\otimes n},
    \qquad
    D_{Q\phase_{\allones}}=\bar\omega S^{\otimes n}+\omega(S^\dagger)^{\otimes n}.
\]
Direct $2\times2$ multiplication gives
\[
    (SH)^3=e^{i\pi/4}I,\qquad (S^\dagger H)^3=e^{-i\pi/4}I,\qquad
    \Tr((SH)^2S^\dagger H)=\Tr((S^\dagger H)^2SH)=0.
\]
Cyclicity of the trace makes every mixed three-factor product vanish. Expanding the eight tensor-power terms leaves only the all-$S$ and all-$S^\dagger$ terms. Thus
\[
    T_3(Q,Q,Q)=\omega^3e^{in\pi/4}+\bar\omega^3e^{-in\pi/4}
    =\frac1{\sqrt2}\cos\!\left(\frac{(n-3)\pi}4\right),
\]
and
\[
    T_3(Q,Q,Q\phase_{\allones})
    =\omega^2\bar\omega e^{in\pi/4}+\bar\omega^2\omega e^{-in\pi/4}
    =\frac1{\sqrt2}\cos\!\left(\frac{(n-1)\pi}4\right).
\]
For even $n$, $\eta=0$ and $\abs{\cos((n-3)\pi/4)}=1/\sqrt2$, so $\abs{c_n}=1/2$. For $n\equiv3\pmod4$, the same formula has absolute cosine $1$, so $\abs{c_n}=1/\sqrt2$. For $n\equiv1\pmod4$, $\eta=\allones$ and $\abs{\cos((n-1)\pi/4)}=1$, again $\abs{c_n}=1/\sqrt2$.
\end{proof}

\subsection{Proof of the matching-certificate coupling}
\label{app:certificate-proof}

\begin{proof}[Proof of \Cref{lem:matching-certificate}]
Condition on $(L,\alpha,\beta,\phi,\psi)$ and the algorithm's randomness, making the algorithm deterministic; average at the end. Set $s:=\abs{\F/M}=\abs L$, and write $\sigma=(-1)^{\tilde\sigma}$, $A(u)=(-1)^{\tilde a_u}$, and $B(v)=(-1)^{\tilde b_v}$ over $\mathbb F_2$, with $(\tilde a,\tilde b)$ uniform on $\mathbb F_2^{2s}$ and independent of $\tilde\sigma$.

Two $f_6$-queries have equal left endpoints if and only if their locations differ by an element of $M$, which happens if and only if their right endpoints agree. The bijection $v(u)=u+[\alpha+\beta]_M$ therefore makes distinct queried edges vertex-disjoint. Every vertex incident to a nonempty set $F$ of distinct queried edges has degree $1$. An odd $F$ has even degree at every unexposed vertex if and only if every edge in $F$ is aligned, and such an $F$ exists if and only if one aligned edge exists. Hence $\Eodd=\Ealign$.

Couple $\tilde\sigma=0$ and $\tilde\sigma=1$ online, stopping immediately before reading the answer to the first query whose location triggers $\Ealign$. Shift-channel answers are fixed by $(L,\alpha,\beta,\phi,\psi)$ and are identical. Queries to $f_2(x_2)$, $f_4(x_4)$, and $f_6(z)$ observe
\[
    \tilde a_{[x_2]_M},\qquad \tilde b_{[x_4]_M},\qquad
    \tilde\sigma+\tilde a_{[z+\alpha]_M}+\tilde b_{[z+\beta]_M}.
\]
Repeated queries repeat answers. Before alignment, each block $(\tilde a_u,\tilde b_{v(u)})$ has at most two of its three forms queried, and any two forms have independent mask components. Distinct blocks use disjoint variables. Thus every fresh coset answer is uniform conditioned on previous answers and has the same law under both signs. Feeding the same fresh sign to both executions keeps transcripts identical through step $t-1$. Both executions therefore issue the same query at step $t$, and $\Ealign$ depends only on query locations and $(L,\alpha,\beta)$, so they trigger it together. Hence $\Prob_+[\Ealign]=\Prob_-[\Ealign]$, and $\TV(\Transcript_+,\Transcript_-)\le\Prob[\Ealign]$.
\end{proof}

\subsection{Proof of the deferred-\texorpdfstring{$f_6$}{f6} alignment bound (Part 1)}
\label{app:posterior-proof}

\begin{proof}[Proof of \Cref{prop:posterior-predictive}, Part 1]
Let $R$ be independent of the ensemble. Under the deferred-$f_6$ hypothesis, all queries to $f_1,\ldots,f_5$ are answered and $Q_6$ is fixed before any $f_6$ answer. Let $Y$ be the ordered shift-channel record and $(Y_2,Y_4)$ the ordered mask-channel record, including step indices, query points, and answers. For fixed $(L,R,\alpha,\beta,\phi,\psi,A,B)$, the record occurs if and only if:
\begin{enumerate}[label=\textup{(\roman*)}]
    \item $\ind_{\mathrm{alg}}(R,Y,Y_2,Y_4)=1$, meaning simulation with $R$ reproduces the recorded queries;
    \item $\ind_Y(\alpha,\beta,\phi,\psi)=1$, meaning the shift oracles agree with $Y$;
    \item $\ind_{Y_2,Y_4}(A,B)=1$, meaning the mask oracles agree with $(Y_2,Y_4)$.
\end{enumerate}
Conditional independence of $(A,B)$ and $(\alpha,\beta,\phi,\psi)$ given $L$ yields
\begin{align*}
    &\Prob[Y,Y_2,Y_4\mid L,R,\alpha,\beta,\phi,\psi]\\
    &\qquad=\ind_{\mathrm{alg}}(R,Y,Y_2,Y_4)\,
      \Prob_{A,B\mid L}[\ind_{Y_2,Y_4}(A,B)=1]\,
      \ind_Y(\alpha,\beta,\phi,\psi).
\end{align*}
Only the last factor depends on $(\alpha,\beta,\phi,\psi)$. Bayes' rule therefore gives, almost surely on admissible records,
\[
    \Prob[(\bar\alpha,\bar\beta)=(a,b)\mid L,R,Y,Y_2,Y_4]
    =\Prob[(\bar\alpha,\bar\beta)=(a,b)\mid L,Y]
    =:\pi_{L,Y}(a,b).
\]
The sets $Q_2,Q_4,Q_6$ are deterministic functions of $(R,Y,Y_2,Y_4)$. Each triple $(x_2,x_4,z)$ aligns only on the single point $([x_2+z]_M,[x_4+z]_M)$, whose conditional probability is at most $\rho(L,Y)$. A union bound over at most $q_2q_4q_6$ triples gives
\[
    \Prob[\Ealign\mid L,R,Y,Y_2,Y_4]
    \le\min\set{1,q_2q_4q_6\rho(L,Y)}.
\]
Taking expectations over $(R,Y_2,Y_4)$ conditioned on $(L,Y)$ proves Part~1.
\end{proof}

\subsection{Shift-channel linear normal form}
\label{app:shift-normal-form}

Fix $L\le\F$ and $M=L^\perp$, and fix the shift-channel query pattern $Q$ of a transcript $Y$. Index query rows by disjoint sets $I_1,I_3,I_5$, one row per query without collapsing repeated points, and write $\abs Q=\abs{I_1}+\abs{I_3}+\abs{I_5}$. Encode signs by $(-1)^b$ and write $\phi(x)=(-1)^{\tilde\phi([x]_L)}$, $\psi(x)=(-1)^{\tilde\psi([x]_L)}$. The response bits satisfy
\[
\begin{aligned}
    y_i&=\alpha\cdot x_i+\tilde\phi([x_i]_L) &&(i\in I_1),\\
    y_i&=(\alpha+\beta)\cdot x_i+\tilde\phi([x_i]_L)+\tilde\psi([x_i]_L) &&(i\in I_3),\\
    y_i&=\beta\cdot x_i+\tilde\psi([x_i]_L) &&(i\in I_5).
\end{aligned}
\]
Collecting responses into $y$ and mask bits into $w_{\mathrm{mask}}$ gives
\[
    y=P(Q)(\alpha,\beta)+M_L(Q)w_{\mathrm{mask}}
    \qquad\text{over }\mathbb F_2.
\]
Let $K_L(Q)$ have rows forming a basis of $\ker(M_L(Q)\T)$, with $r:=\dim\ker(M_L(Q)\T)=\abs Q-\rankF(M_L(Q))$.

For $\lambda\in\ker(M_L(Q)\T)$, the coefficient of $\tilde\phi(C)$ on any coset $C$ is an even sum of selected rows from $I_1\sqcup I_3$. Since an even sum of vectors from one $L$-coset lies in $L$, the coefficient $u_\lambda:=\sum_{i\in I_1\sqcup I_3}\lambda_ix_i$ of $\alpha$ lies in $L$. Similarly, the coefficient $v_\lambda$ of $\beta$ lies in $L$. Thus $\lambda\T P(Q)(\alpha,\beta)$ depends only on $(\bar\alpha,\bar\beta)$, and $K_L(Q)P(Q)$ factors through a map $\overline P(Q):(\F/M)^2\to\mathbb F_2^r$. Define
\[
    \affrank(L,Q):=\rankF(\overline P(Q))\le r,
\]
with $\affrank(L,Q)=0$ when $r=0$. Every realized transcript satisfies the consistent system
\begin{equation}
    \overline P(Q)(\bar\alpha,\bar\beta)=K_L(Q)y.
    \label{eq:shift-eliminated}
\end{equation}

\subsection{Affine posterior peak (Part 2)}
\label{app:posterior-rank}

\begin{proof}[Proof of \Cref{prop:posterior-predictive}, Part 2]
Fix $(L,Y)$ in the uniform-shift ensemble. By the factorization in \Cref{app:posterior-proof}, the conditional probability of $(\alpha,\beta)$ is proportional to the number of mask vectors satisfying $M_L(Q)w_{\mathrm{mask}}=y+P(Q)(\alpha,\beta)$. The system is consistent exactly when $K_L(Q)(y+P(Q)(\alpha,\beta))=0$, and every consistent right-hand side has the same number of solutions. The prior on $(\bar\alpha,\bar\beta)$ is uniform on $(\F/M)^2$, of size $s^2$. Therefore the posterior is uniform on the affine solution set of \eqref{eq:shift-eliminated}. Since that system is consistent and $\overline P(Q)$ has rank $\affrank(L,Y)$, the solution set has cardinality $s^2/2^{\affrank(L,Y)}$. Hence $\rho(L,Y)=2^{\affrank(L,Y)}/s^2$, and Part~1 gives
\[
    \Prob[\Ealign\mid L,Y]
    \le\min\!\left\{1,\frac{q_2q_4q_6\,2^{\affrank(L,Y)}}{s^2}\right\}.
\]
\end{proof}

\section{The adaptive upper bound (full algorithm)}
\label{app:ensemble-upper}

Throughout this appendix, fix $n\ge1$, $N:=2^n$, a subspace $L\le\F$ with $M:=L^\perp$, and sizes $s:=\abs L=\abs{\F/M}$ and $t:=\abs M=\abs{\F/L}=N/s$, so $st=N$. For a Boolean function $g$, let $\Per(g):=\set{v\in\F:g(x+v)=g(x)\ \forall x}$ and define
\[
    \Meff:=\Per(f_2)\cap\Per(f_4)=\Per(A)\cap\Per(B),
    \qquad \Leff:=\Meff^\perp.
\]
Because $A,B$ are $M$-periodic, $M\subseteq\Meff$ and $\Leff\subseteq L$.

\subsection{Exact decoding and period-testing lemmas}
\label{app:tests}

\begin{lemma}[Exact decoding from a sandwiched period subspace or coset pair]
\label{lem:eff-decode}
Let $(f_1,\ldots,f_6)$ be any realization of \eqref{eq:t6-family}.
\begin{itemize}
    \item \textbf{Part 1 (Sandwiched period subspace).} If a known subspace $M'\le\F$ satisfies $M\subseteq M'\subseteq\Meff$, equivalently $\Leff\subseteq L'\subseteq L$ for $L':=(M')^\perp$, then $\sigma$ is computed exactly with at most $2n+5$ further queries.
    \item \textbf{Part 2 (Certified coset pair).} If $M'\subseteq\Meff$ and a pair $(w,w')\in(\alpha+M')\times(\beta+M')$ is known, then $\sigma$ is computed exactly with $3$ queries.
\end{itemize}
\end{lemma}

\begin{proof}
For Part~1, let $d:=\dim L'\le n$. Since $\Leff\subseteq L'\subseteq L$, $\phi,\psi$ are constant on $L'$-cosets. Choose a basis $\ell_1,\ldots,\ell_d$ of $L'$ (empty if $d=0$). Query $f_1(0),f_5(0)$ and $f_1(\ell_i),f_5(\ell_i)$. Then
\[
    f_1(0)f_1(\ell_i)=(-1)^{\alpha\cdot\ell_i},\qquad
    f_5(0)f_5(\ell_i)=(-1)^{\beta\cdot\ell_i}.
\]
Solving these consistent systems gives $\alpha_0,\beta_0$ with $\alpha_0+\alpha\in(L')^\perp=M'$ and $\beta_0+\beta\in M'$. Querying $f_2(\alpha_0),f_4(\beta_0),f_6(0)$ gives
\[
    f_2(\alpha_0)f_4(\beta_0)f_6(0)
    =A(\alpha_0)B(\beta_0)\sigma A(\alpha)B(\beta)=\sigma,
\]
because $\alpha_0+\alpha\in\Per(A)$ and $\beta_0+\beta\in\Per(B)$. The total is $2d+2+3\le2n+5$. For Part~2, $w+\alpha\in M'\subseteq\Per(A)$ and $w'+\beta\in M'\subseteq\Per(B)$, so $f_2(w)f_4(w')f_6(0)=\sigma$.
\end{proof}

Here we introduce the heavy Fourier-coefficient recovery theorem of \citet{GoldreichLevin1989}.

\begin{fact}[Goldreich--Levin heavy-coefficient recovery; \citealp{GoldreichLevin1989}]
\label{fac:goldreich-levin}
For every $n\ge1$, Boolean function $g\colon\F\to\set{-1,+1}$, threshold $\tau\in(0,1]$, and $\delta\in(0,1)$, there is a randomized query algorithm $\textsc{GoldreichLevin}(g,\tau,\delta)$ making $\operatorname{poly}(n,1/\tau,\log(1/\delta))$ queries and, with probability at least $1-\delta$, outputting a list $\Lambda\subseteq\F$ of size $\operatorname{poly}(n,1/\tau,\log(1/\delta))$ containing every $\xi$ with $\abs{\widehat g(\xi)}\ge\tau$.
\end{fact}

The three tests are
\[
\begin{aligned}
    T_A(v)&:\quad x\sim\Unif(\F),\quad\text{accept iff }f_2(x)=f_2(x+v),\\
    T_B(v)&:\quad x\sim\Unif(\F),\quad\text{accept iff }f_4(x)=f_4(x+v),\\
    D_\ell&:\quad x,x'\sim\Unif(\F)\text{ independently},\quad
      \text{accept iff }f_1(x)f_1(x+\ell)=f_1(x')f_1(x'+\ell).
\end{aligned}
\]
The first two cost two queries each; $D_\ell$ costs four.

\begin{lemma}[Per-trial rejection probabilities]
\label{lem:test-reject}
Let $(f_1,\ldots,f_6)$ be drawn from the uniform-shift ensemble.
\begin{itemize}
    \item \textbf{Part 1.} Every $v\in\Meff$ passes $T_A,T_B$ with probability $1$. Every $v\notin\Meff$ is rejected by the pair with probability at least $2/s$ per trial.
    \item \textbf{Part 2.} Every $\ell\in L$ passes $D_\ell$ with probability $1$. For fixed $\ell\notin L$, write $q_\ell(x):=\phi(x)\phi(x+\ell)$ and $p_+(\ell):=\Prob_x[q_\ell(x)=+1]$. With probability at least $1-2e^{-t/16}$ over $\phi$, $p_+(\ell)\in[1/4,3/4]$, and then $D_\ell$ rejects with probability at least $3/8>1/4$.
\end{itemize}
\end{lemma}

\begin{proof}
For Part~1, if $v\notin\Meff$, at least one mask, say $A$, has $A(x_0)\ne A(x_0+v)$ for some $x_0$. Since $v\notin M$, the two $M$-cosets $x_0+M$ and $x_0+v+M$ are disjoint and each has $N/s$ elements. By $M$-periodicity, every $x$ in either coset has $A(x)\ne A(x+v)$, so rejection probability is at least $2/s$.

For Part~2, $\ell\in L$ makes $f_1(x)f_1(x+\ell)=(-1)^{\alpha\cdot\ell}$ independent of $x$. If $\ell\notin L$, translation by $[\ell]_L$ partitions the $t$ cosets into $t/2$ pairs. On each pair, $q_\ell$ takes an independent uniform sign $Z_i$. Thus $p_+(\ell)$ is the average of $t/2$ independent Bernoulli$(1/2)$ variables. Hoeffding gives
\[
    \Prob_\phi[\abs{p_+(\ell)-1/2}>1/4]
    \le2e^{-2(t/2)(1/4)^2}=2e^{-t/16}.
\]
If $p_+(\ell)\in[1/4,3/4]$, two independent draws give equal signs with probability $p_+^2+(1-p_+)^2\le5/8$, so rejection probability is at least $3/8$.
\end{proof}

\begin{lemma}[Uniform rejection events for large $s$ or large $t$]
\label{lem:good-event}
\begin{itemize}
    \item \textbf{Part 1 (Large $s$).} If $s\ge32n$, with probability at least $1-(s-1)e^{-s/16}\ge1-32ne^{-2n}$ over $A$, every $v\notin M$ is rejected by $T_A(v)$ with probability at least $1/4$ per trial. On this event $\Per(A)=M=\Meff$.
    \item \textbf{Part 2 (Large $t$).} If $t\ge32n$, with probability at least $1-2(t-1)e^{-t/16}\ge1-64ne^{-2n}$ over $\phi$, every $\ell\notin L$ has $p_+(\ell)\in[1/4,3/4]$ and $D_\ell$ rejects with probability at least $3/8$.
\end{itemize}
\end{lemma}

\begin{proof}
For Part~1, fix $v\notin M$. Translation by $[v]_M$ partitions the $s$ cosets into $s/2$ pairs. Each pair independently disagrees with probability $1/2$, and both cosets of a disagreeing pair reject $T_A(v)$. Hence rejection probability is the average of $s/2$ independent Bernoulli$(1/2)$ variables. It is below $1/4$ with probability at most $e^{-2(s/2)(1/4)^2}=e^{-s/16}$. A union bound over $s-1$ nonzero cosets gives $(s-1)e^{-s/16}\le32ne^{-2n}$ for $s\ge32n$. On this event no $v\notin M$ belongs to $\Per(A)$, so $\Per(A)=M=\Meff$.

For Part~2, apply a union bound over the $t-1$ nonzero cosets to Part~2 of \Cref{lem:test-reject}. The same monotonicity gives at most $64ne^{-2n}$ for $t\ge32n$.
\end{proof}

\subsection{Certified four-branch adaptive algorithm}
\label{app:branches}

\Cref{alg:adaptive-cs} does not know $(s,t)$. Branch~4 runs first and is self-certifying. If it does not return, $t>256n^2$ holds except with probability $2^{-\Omega(n)}$, so the differential certifier applies. Candidate subspaces are decoded only after both inclusions $M'\subseteq\Meff$ and $M\subseteq M'$ are tested.

\begin{algorithm}[!htbp]
\caption{Certified four-branch adaptive recovery of $\sigma$.}
\label{alg:adaptive-cs}
\begin{paperalgorithmic}[1]
\Require Oracle access to $(f_1,\ldots,f_6)$ on $\F$ from the uniform-shift ensemble; $n\ge1$.
\Ensure $\hat\sigma\in\set{-1,+1}$ or \textsc{fail}.
\State \textbf{Branch 4:} Set $\tau\gets1/(16n)$, $\delta_{\mathrm{GL}}\gets2^{-n}$, $K_{\mathrm{est}}\gets8192n^3$.
\State Run $\Lambda_1\gets\textsc{GoldreichLevin}(f_1,\tau,\delta_{\mathrm{GL}})$ and $\Lambda_5\gets\textsc{GoldreichLevin}(f_5,\tau,\delta_{\mathrm{GL}})$.
\State For each listed $\xi$, estimate the corresponding Fourier coefficient by $K_{\mathrm{est}}$ independent samples. Retain it iff the estimate has magnitude $\ge\tau/2$.
\If{both filtered lists are nonempty}
    \State Choose $w,w'$ from them and \Return $f_2(w)f_4(w')f_6(0)$.
\EndIf
\State \textbf{Branch 1:} Set $S_1\gets\emptyset$ and $r\gets4n$.
\For{$b\in\set{2,4,\ldots,2^{\lceil n/4\rceil+1}}$ and trial $i=1,\ldots,16n$}
    \State Draw uniform $u_1,\ldots,u_r,v_1,\ldots,v_b$ and query $f_2(u_a+v_j)$ for all $a,j$.
    \State For lexicographic pairs with equal signatures $(f_2(u_a+v_j))_{a=1}^r$, testing at most $n^5$ candidates, test $\delta=v_j+v_{j'}$ by $16n$ fresh $T_A(\delta)$ trials. On the first pass, add $\delta$ to $S_1$ and exit the trial.
\EndFor
\State Set $M_1\gets\operatorname{span}(S_1)$.
\State \textbf{Branch 2:} Set $S_2\gets\emptyset$ and $r\gets4n$.
\For{$b\in\set{2,4,\ldots,2^{\lceil n/4\rceil+1}}$ and trial $i=1,\ldots,16n$}
    \State Draw uniform $u_0,u_1,\ldots,u_r,v_1,\ldots,v_b$ and query $f_1(u_a+v_j)$ for all $a,j$.
    \State Form $Z_j\gets(f_1(u_0+v_j)f_1(u_a+v_j))_{a=1}^r$. For equal-signature pairs, testing at most $n^5$ candidates, test $\delta=v_j+v_{j'}$ by $16n$ fresh $D_\delta$ trials. On the first pass, add $\delta$ to $S_2$ and exit the trial.
\EndFor
\State Set $M_2\gets(\operatorname{span}(S_2))^\perp$.
\State \textbf{Branch 3:} Set $S_3\gets\emptyset$.
\For{$i=1,\ldots,n^4$}
    \State Draw uniform nonzero $v$ and run $64n^3$ fresh $(T_A(v),T_B(v))$ trials. If all pass, add $v$ to $S_3$.
\EndFor
\State Set $M_3\gets\operatorname{span}(S_3)$.
\State \textbf{Certification and decoding:}
\For{$M'\in\set{M_1,M_2,M_3}$}
    \State Compute bases of $M'$ and $L':=(M')^\perp$.
    \If{every basis vector of $M'$ passes $64n^3$ fresh $(T_A,T_B)$ trials and every basis vector of $L'$ passes $16n$ fresh $D_\ell$ trials}
        \State \Return exact $\sigma$ via Part~1 of \Cref{lem:eff-decode}.
    \EndIf
\EndFor
\State \Return \textsc{fail}.
\end{paperalgorithmic}
\end{algorithm}

\subsection{Complete proof of \Cref{prop:ensemble-upper}}
\label{app:upper-proof}

\begin{proof}[Proof of \Cref{prop:ensemble-upper}]
We show that \Cref{alg:adaptive-cs} makes $n^{O(1)}N^{1/4}$ queries and outputs $\sigma$ with probability $1-2^{-\Omega(n)}$ for every $L$. Assume $n\ge512$; smaller $n$ are absorbed in $o(1)$.

\paragraph{Step 1: Branch 4 is sound for all $(s,t)$ and complete for $t\le256n^2$.}
By \Cref{app:t6-proof}, $\supp(\widehat{f_1})\subseteq\alpha+M$ and $\supp(\widehat{f_5})\subseteq\beta+M$. For any listed $\xi$, each sample lies in $[-1,1]$. With $K_{\mathrm{est}}=8192n^3=32n/\tau^2$, Hoeffding gives
\[
    \Prob[\abs{\tilde\mu(\xi)-\widehat g(\xi)}\ge\tau/4]
    \le2e^{-K_{\mathrm{est}}\tau^2/32}=2e^{-n}.
\]
By \Cref{fac:goldreich-levin}, the lists have polynomial size, so a union bound makes all estimates accurate within $\tau/4$ with probability $1-2^{-\Omega(n)}$. Every retained frequency then has true magnitude at least $\tau/4>0$ and lies in the correct coset. If Branch~4 returns, Part~2 of \Cref{lem:eff-decode} outputs $\sigma$ exactly.

If $t\le256n^2$, Parseval over a support of size $t$ gives a coefficient of magnitude at least $t^{-1/2}\ge1/(16n)=\tau$ for each of $f_1,f_5$. \Cref{fac:goldreich-levin} finds both with probability at least $1-2\cdot2^{-n}$, and the filter retains them. Thus Branch~4 succeeds with probability $1-2^{-\Omega(n)}$.

\paragraph{Step 2: Soundness of the universal certifier when $t>256n^2$.}
Assume Branch~4 did not return. Then $t>256n^2$ except with probability $2^{-\Omega(n)}$, so the large-$t$ event of \Cref{lem:good-event} holds with probability $1-64ne^{-2n}$. If $s\le64n^2$, \Cref{lem:test-reject} gives rejection probability at least $2/s\ge1/(32n^2)$ for every $v\notin\Meff$. If $s>64n^2$, the large-$s$ event holds with probability $1-32ne^{-2n}$ and gives rejection probability at least $1/4$.

For a candidate $M'$, if $M'\not\subseteq\Meff$, some basis vector $m_a$ lies outside $\Meff$ and passes $64n^3$ tests with probability at most $(1-1/(32n^2))^{64n^3}\le e^{-2n}$. If $M\not\subseteq M'$, then $L'\not\subseteq L$, so some basis vector $\ell_a$ lies outside $L$ and passes $16n$ differential tests with probability at most $(3/4)^{16n}<e^{-4n}$. Conversely, if $M\subseteq M'\subseteq\Meff$, every basis vector passes with probability $1$, and \Cref{lem:eff-decode} is exact. A union bound over three candidates bounds false certification by $2^{-\Omega(n)}$.

\paragraph{Step 3: Branch 3 recovers $\Meff$ when $s\le64n^2$ and $t>256n^2$.}
Each $v\notin\Meff$ survives $64n^3$ tests with probability at most $(1-2/s)^{64n^3}\le e^{-2n}$, so $M_3\subseteq\Meff$ with probability at least $1-n^4e^{-2n}$. Since $\abs{\Meff}\ge t=N/s>256n^2$, a uniform nonzero draw lands in $\Meff\setminus\set0$ with probability
\[
    p_3=\frac{\abs{\Meff}-1}{N-1}\ge\frac{N/s-1}{N-1}\ge\frac1{2s}\ge\frac1{128n^2}.
\]
The expected count is at least $n^2/128$. Chernoff gives at least $n^2/256\ge2n$ such draws with probability $1-e^{-\Omega(n^2)}$. Conditional on at least $2n$ independent uniform nonzero vectors in $\Meff$, they fail to span it only if all lie in one hyperplane. Each hyperplane contains less than half the nonzero vectors, and there are fewer than $2^n$ hyperplanes, so failure probability is below $2^{n-2n}=2^{-n}$. Hence $M_3=\Meff$ with probability $1-2^{-\Omega(n)}$.

\paragraph{Step 4: Branch 1 recovers $M$ when $32n\le s\le N^{1/2}$.}
Condition on the large-$s$ event. Across all scales and trials, at most $16n^7$ candidates are tested. Every $\delta\notin M$ passes $16n$ tests with probability at most $(3/4)^{16n}$, so $M_1\subseteq M$ with probability $1-2^{-\Omega(n)}$.

Because $32n\le s\le N^{1/2}$,
\[
    2<2\sqrt s\le2\sqrt{N^{1/2}}=2N^{1/4}\le2^{\lceil n/4\rceil+1},
\]
so some power-of-two scale $b^*\in[2\sqrt s,4\sqrt s)$ exists. For a false pair with $\delta\notin M$, the shifted points $u_a+v_j$ are i.i.d.\ uniform, and each comparison agrees with probability at most $3/4$ on the large-$s$ event. Thus a false signature collision has probability at most $(3/4)^{4n}$. Since $\binom{b^*}2<8s\le8\cdot2^{n/2}$, the expected number of false collisions per trial is at most $8\cdot2^{n/2}(3/4)^{4n}=2^{-\Omega(n)}$. By Markov and a union bound over $16n$ trials, no false collision occurs at scale $b^*$ with high probability.

For true collisions, decompose $v_j=c_j+m_j$ with $c_j$ a uniform coset representative and $m_j$ uniform in $M$, independently. The probability that all $c_j$ are distinct is at most $\exp(-\binom{b^*}2/s)$. Since $\binom{b^*}2/s\ge2-1/\sqrt s\ge1$, a true pair occurs with probability at least $1-e^{-1}>0.63$. Signatures and $T_A$ outcomes depend only on the cosets, while $\delta=m_j+m_{j'}$ is uniform in $M$ and independent across trials. Every true pair passes all tests. Over $16n$ trials, at least $4n$ uniform elements of $M$ are collected with probability $1-2^{-\Omega(n)}$. A hyperplane union bound shows they span $M$ with probability at least $1-2^{-3n}$. Combining both inclusions gives $M_1=M$.

\paragraph{Step 5: Branch 2 recovers $M$ when $32n\le t\le N^{1/2}$.}
Condition on the large-$t$ event. The signature
\[
    f_1(u_0+v_j)f_1(u_a+v_j)
    =(-1)^{\alpha\cdot(u_0+u_a)}\phi(u_0+v_j)\phi(u_a+v_j)
\]
depends on $v_j$ only through $[v_j]_L$. For $\delta\notin L$, conditioning on $(v_j,v_{j'},u_0)$ fixes the right-hand sign, while $u_a+v_j$ remains i.i.d.\ uniform. Each coordinate matches with probability at most $3/4$, giving false collision probability at most $(3/4)^{4n}$. Every $\delta\notin L$ passes $16n$ differential tests with probability at most $(3/4)^{16n}$. Repeating Step~4 with $L$ in place of $M$ and $t$ in place of $s$ gives $\operatorname{span}(S_2)=L$ with probability $1-2^{-\Omega(n)}$, hence $M_2=L^\perp=M$.

\paragraph{Step 6: Coverage and query complexity.}
If $t\le256n^2$, Branch~4 succeeds. If $t>256n^2$ and $s\le64n^2$, Branch~3 succeeds. If $t>256n^2$ and $64n^2<s\le N^{1/2}$, Branch~1 succeeds. If $t>256n^2$ and $s>N^{1/2}$, then $32n<256n^2<t=N/s<N^{1/2}$, so Branch~2 succeeds. \Cref{tab:branches,fig:regimes} summarize the regimes.

Branch~4 costs $n^{O(1)}$ queries. Each birthday branch costs $\sum_{j=1}^{\lceil n/4\rceil+1}16n(4n+1)2^j=O(n^2N^{1/4})$ signature queries and at most $(\lceil n/4\rceil+1)16n\cdot n^5\cdot(16n\cdot4)=O(n^8)$ verification queries. Branch~3 costs $O(n^7)$. Certifying three subspaces costs $O(n^4)$, and decoding costs at most $2n+5$. The total is $O(n^2N^{1/4}+n^8)+n^{O(1)}=n^{O(1)}N^{1/4}$.
\end{proof}

\subsection{Regime coverage summary}
\label{app:regime-table}

The regime diagram is \Cref{fig:regimes} in the main body; the following table records the same coverage cases.

\begin{table}[t]
\centering
{\scriptsize
\setlength{\tabcolsep}{4pt}
\begin{tabular}{@{}>{\raggedright\arraybackslash}p{0.20\linewidth}>{\raggedright\arraybackslash}p{0.28\linewidth}>{\raggedright\arraybackslash}p{0.18\linewidth}>{\raggedright\arraybackslash}p{0.28\linewidth}@{}}
\toprule
Regime ($st=N$) & Branch in \Cref{alg:adaptive-cs} & Query cost & Certified object \\
\midrule
$t\le256n^2$ & Branch~4 (Goldreich--Levin + filter) & $n^{O(1)}$ & coset pair $(w,w')\in(\alpha+M)\times(\beta+M)$ \\
\addlinespace
$t>256n^2,\ s\le64n^2$ & Branch~3 (direct random periods) & $O(n^7)$ & $\Meff$ (verified $M\subseteq\Meff$) \\
\addlinespace
$32n\le s\le N^{1/2}$ & Branch~1 (birthday on $f_2$) & $O(n^2N^{1/4}+n^8)$ & $M$ exactly \\
\addlinespace
$32n\le t\le N^{1/2}$ & Branch~2 (differential birthday on $f_1$) & $O(n^2N^{1/4}+n^8)$ & $L$ exactly ($M_2=L^\perp=M$) \\
\bottomrule
\end{tabular}}
\caption{Regime coverage of \Cref{alg:adaptive-cs}. Every $(s,t)$ with $st=N$ meets at least one row; once certified, \Cref{lem:eff-decode} recovers $\sigma$ exactly.}
\label{tab:branches}
\end{table}

\end{document}